
\input tables.tex
\input phyzzx.tex

\def\dashes{($-~-~-~-$)}
\def\dots{($\cdot~\cdot~\cdot~\cdot~\cdot~\cdot\,$)}

\def\dotdash{($\cdot~-~\cdot~-$)}

\def\hpm{H^{\pm}}
\def\mhpm{m_{\hpm}}
\def\mhp{m_{\hp}}
\def\hp{H^+}
\def\hm{H^-}
\def\aachen{{\it Proceedings of the Large Hadron
Collider Workshop}, edited by G. Jarlskog and D. Rein, Aachen (1990),
CERN 90-10, ECFA 90-133}
\def\tauptaum{\tau^+\tau^-}

\def\chitil{\widetilde\chi}
\def\cnone{\chitil^0_1}
\def\hsm{\phi^0}
\def\mhsm{m_{\hsm}}
\def\half{\ifmath{{\textstyle{1 \over 2}}}}
\def\h{h}
\def\mh{m_{\h}}
\def\tanb{\tan\beta}

\def\sinb{\sin\beta}

\def\chitil{\widetilde\chi}

\def\cnone{\chitil^0_1}

\def\wpm{W^{\pm}}
\def\wp{W^+}
\def\wm{W^-}
\def\hl{h^0}
\def\mhl{m_{\hl}}
\def\hh{H^0}
\def\mhh{m_{\hh}}
\def\hpm{H^{\pm}}
\def\mhpm{m_{\hpm}}
\def\mt{m_t}
\def\mb{m_b}

\def\hl{h^0}
\def\hh{H^0}
\def\ha{A^0}
\def\mhl{m_{\hl}}
\def\mhh{m_{\hh}}
\def\mha{m_{\ha}}
\def\eps{\epsilon}

\def\gam{\gamma}
\def\gam{\gamma}
\def\ptr{p_T}
\def\mtwo{M_{jj}}
\def\mthree{M_{bjj}}
\def\mfour{M_{bbjj}}
\def\mmissl{M_{miss-\ell}}

\def\delmw{\Delta\mw}
\def\delmt{\Delta\mt}
\def\gam{\gamma}
\def\ebtag{e_{b-tag}}
\def\emistag{e_{mis-tag}}
\def\lamnew{\Lambda_{\rm New Physics}}

\def\stop{{\wtilde t}}
\def\mstop{m_{\stop}}

\def\ptr{p_T}
\def\etmiss{E_T^{\,miss}}
\def\etell{E_T^{\,\ell}}
\def\ptmiss{\vec p_T^{\,\,miss}}
\def\ptell{\vec p_T^{\,\,\ell}}
\def\mmissl{M_{miss-\ell}}

\def\gev{~{\rm GeV}}
\def\tev{~{\rm TeV}}
\def\pbi{~{\rm pb}^{-1}}
\def\fbi{~{\rm fb}^{-1}}

\def\pb{~{\rm pb}}
\def\hawaii{{\it Proceedings of the 2nd International Workshop on
``Physics and Experiments with Linear $\epem$ Colliders''}, ed. F. Harris,
Waikoloa, HI, April 26-30, 1993}
\def\smv{{\it Proceedings of the 1990 DPF Summer Study on
High Energy Physics: ``Research Directions for the Decade''},
editor E. Berger, Snowmass (1990)}
\def\perspectives{{\it Perspectives on Higgs Physics}, ed. G. Kane, World
Scientific Publishing, Singapore (1992)}
\def\madisonargonne{{\it Proceedings of the ``Workshop on Physics at Current
Accelerators and the Supercollider''}, eds. J. Hewett, A. White, and
D. Zeppenfeld, Argonne National Laboratory, 2-5 June (1993)}

\def\prdj#1{{\it Phys. Rev.} {\bf D{#1}}}
\def\npbj#1{{\it Nucl. Phys.} {\bf B{#1}}}
\def\prlj#1{{\it Phys. Rev. Lett.} {\bf {#1}}}
\def\plbj#1{{\it Phys. Lett.} {\bf B{#1}}}
\def\zpcj#1{{\it Z. Phys.} {\bf C{#1}}}

\def\mt{m_t}
\def\wp{W^+}
\def\wm{W^-}
\def\rta{\rightarrow}
\def\tanb{\tan\beta}
\def\sinb{\sin\beta}

\def\lplm{\ell^+\ell^-}
\def\lplp{\ell^+\ell^+}

\def\hn{h}
\def\hsm{\phi^0}
\def\mhn{m_\hn}
\def\mhsm{m_{\hsm}}
\def\nsd{N_{SD}}

\def\mw{m_W}
\def\mz{m_Z}
\def\anti{\overline}

\def\ifmath#1{\relax\ifmmode #1\else $#1$\fi}
\def\half{\ifmath{{\textstyle{1 \over 2}}}}

\def\3quarter{{\textstyle{3 \over 4}}}

\def\eps{\epsilon}
\def\ebtag{e_{b-tag}}
\def\emisid{e_{mis-id}}
\def\tth{t\anti t \hn}
\def\ttz{t\anti t Z}
\def\ttglnu{t\anti t g-\ell\nu}
\def\ttbdecay{t\anti t g-b\ell\nu}

\input phyzzx
\Pubnum={$\caps UCD-93-32$\cr}
\date{September, 1993}

\titlepage
\vskip 0.75in
\baselineskip 14pt
\hsize=6in
\vsize=8.5in
\centerline{\bf PROGRESS IN SSC HIGGS PHYSICS:}
\centerline{{\bf REPORT OF THE HIGGS WORKING GROUP}
\foot{To appear in \madisonargonne.}
}
\vskip .5in
\centerline{ J.F. Gunion$^{(a)}$ and S. Geer$^{(b)}$}
\vskip .075in
\centerline{\it a) Davis Institute for High Energy Physics,
Dept. of Physics, U.C. Davis, Davis, CA 95616}
\vskip .075in
\centerline{\it b) Fermi National Accelerator Laboratory, Batavia, IL 60510}
\vskip .075in
\noindent Subgroup convenors: Light Higgs --- L. Orr; Heavy Higgs --- T.
Han
\vskip .075in
\noindent Working Group Members and Contributors:
P. Agrawal, J. Bagger, B. Bailey, V. Barger,
G. Belanger, R. Blair,  D. Chao,
K. Cheung, N. Deshpande, A. Djouadi, T. Garavaglia, D. Graudenz, J. Hewett,
R. Kauffman, W. Kwong, W. Marciano, D. Morris, J. Qiu, T. Rizzo,
W. Schaffer, M. Shaw, A. Stange, D. Summers, S. Willenbrock, D.
Wu, C.P. Yuan, D. Zeppenfeld

\vskip .075in
\centerline{\bf Abstract}
\vskip .075in
\centerline{\Tenpoint\baselineskip=12pt
\vbox{\hsize=12.4cm
\noindent I review new developments in Higgs physics and electroweak
symmetry breaking that have resulted from the Madison--Argonne
workshops on SSC physics.
}}

\vskip .15in
\noindent{\bf 1. Introduction}
\vskip .075in

The most fundamental mission of the SSC will be to reveal the nature and
source of electroweak symmetry breaking (EWSB).  Among the many models
that have been developed for the EWSB sector, the minimal Standard Model
(MSM) and the minimal supersymmetric extension of the Standard Model (MSSM)
are certainly the simplest examples of two quite different classes
of theory, and it is upon these that our working group
focused.  In any case, the issues that arise in developing techniques for
detecting the single Higgs boson
($\hsm$) of the MSM or the family of Higgs bosons that emerge in the MSSM
are representative of those that must be faced in any model.

In the case of the SM $\hsm$, it is well known that discovery is most
challenging if the Higgs is either light or very heavy. Much of the
working group focus was on the so-called `intermediate mass' region,
$80\lsim\mhsm\lsim 2\mz$. Perhaps it is
worth reviewing the motivation for such a focus.
First, all current lattice and related investigations
appear to require that $\mhsm\lsim 650$ GeV if the scale of new
physics $\lamnew$ is to lie above $\mhsm$.
\Ref\hhg{For a review see
J.F. Gunion, H.E. Haber, G.L. Kane, and S. Dawson, {\it The Higgs Hunter's
Guide}, Addison-Wesley, Redwood City, CA (1990).}\
If $\lamnew$ is as
heavy as $\sim 10^{15} GeV$, then $\mhsm\lsim 200$ GeV is required
(by the renormalization group equations
\foot{Of course, these same renormalization group equations
have difficulty reproducing the low-energy value
of $\sin^2\theta_W$ in the simplest SU(5) grand unification scheme.}) in
order that the theory remain perturbative up to the scale $\lamnew$.
Finally, in order that the quartic coupling of the Higgs sector
not be driven to negative values (implying instability of the potential)
by the large Yukawa coupling associated with the heavy top quark,
it is necessary that $\mhsm$ lie above an $\mt$ and $\lamnew$ dependent
lower bound. For $\mt=150$ and $\lamnew\sim 10^{15}$ GeV, for instance,
$\mhsm\gsim 100\gev$ is required in the context of perturbatively computed
renormalization group equations. The lower bound decreases with decreasing
$\lamnew$ and/or $\mt$. Nonetheless, it is entirely reasonable that $\mhsm$
should lie in a range that is somewhat above the current upper limit of
$\sim 60\gev$ set by LEP-I, and quite possibly the $\hsm$ will turn out
to be too heavy to be found at LEP-II (which will probe up to $\mhsm\sim
80-90\gev$ for $\sqrt s\sim 190-200\gev$).

Several techniques for detecting a light $\hsm$ have been shown to be
very promising. For $80\lsim \mhsm\lsim
140\gev$, the $\hsm$ should be visible in $gg\rta \hsm\rta\gam\gam$ at those
detectors with excellent ($\sim 1\%$) $\gam\gam$ mass resolution,
and in the $W\hsm+
t\anti t\hsm\rta \ell\gam\gam X$ mode at all detectors
(with at least $\sim 3\%$ mass resolution). These channels were
critically examined at the workshops, and technical improvements
and uncertainties/concerns will be be reviewed here and in associated
contributed papers.
For $130\lsim\mhsm\lsim 700-800\gev$, an extremely clean signal will be
available in $gg\rta \hsm\rta ZZ^{(*)}\rta 4\ell$.  The viability of the
$4\ell$ channel seems unquestionable in the indicated mass range.
For $\mhsm$ in the 500-800 GeV range, $gg\rta \hsm \rta ZZ\rta \lplm
\nu\anti\nu$ also provides a viable signal; a possible means for extending
this signal to $\mhsm\gsim 800\gev$ was examined during the workshops,
and will be outlined. Missing from the above list is any means for
detecting a light $\hsm$ in its primary $b\anti b$ decay mode.
During the workshops, it was shown
that expected $b$-tagging efficiency and purity should be adequate
to isolate an $\hsm$ signal for $\mhsm\lsim 110-120 \gev$ at the SSC/LHC
in $t\anti t \hsm$ production followed by $\hsm\rta b\anti b$ decay.
In addition, the possibility of detecting
$W\hsm$ production followed by $\hsm\rta b\anti b$ at the Tevatron
(for $\mhsm$ below about 90 GeV) was explored. These possibilities
will also be reviewed.

At the very high end of the mass scale, there has been recent work
showing that we should be able to explore EWSB even if the $\hsm$ is
heavier than 1 TeV
(or non-existent) and the interactions of (longitudinally polarized) $W$
($W\equiv \wpm,Z$) bosons become non-perturbative at high energies.
This work will not be covered in detail in this report. It is reviewed in
these proceedings in the contributions by
J. Bagger, K. Cheung, and D. Morris. A few brief remarks will appear later.

Of course, even in the context of perturbative theories
containing elementary Higgs bosons, the MSM need not be nature's
choice. Many generalizations have been  discussed,\refmark\hhg\
including extensions
of the Higgs sector only, extensions of both the gauge and Higgs sectors,
and supersymmetric generalizations of all these types of models.
Supersymmetric generalizations are particularly attractive in the
perturbative context in that they {\it require} the presence of elementary
spin-zero Higgs fields and solve the well-known naturalness
and hierarchy problems.
Thus, they provide an enormously attractive
theoretical framework in which elementary Higgs bosons must exist. Further,
in supersymmetric models, there is always one (or more) light Higgs
boson(s)  with coupling(s) to the $WW$ channels such that $WW$ scattering
remains perturbative at all energies.

The most thoroughly investigated
model is the Minimal Supersymmetric Model (MSSM) in which the Higgs sector
contains two Higgs-doublet fields (the minimum number required in the
supersymmetric context), but there is no extension of the gauge or matter
sectors other than the introduction of the supersymmetric partner
states. For simple boundary conditions, grand unification in the MSSM
context of perturbative renormalization/evolution equations yields highly
satisfactory values for $\sin^2\theta_W$ and other precisely
measured electroweak parameters. In
addition, a common GUT-scale Yukawa coupling yields fermion mass ratios,
\eg\ $\mb/m_\tau$ that tend to be in close agreement with experiment.
The MSSM also provides a good candidate for dark matter (the lightest
neutralino), and the predicted GUT scale adequately suppresses proton decay.

The Higgs sector of the MSSM is a highly constrained two-doublet
Higgs model.  Two doublets are required by the basic structure of
supersymmetry which makes it impossible to use a single Higgs superfield
and its complex conjugate simultaneously in the
superpotential construction that is responsible for fermion masses.
(Recall that in the MSM the Higgs field yields the down quark masses,
while its complex conjugate appears in the Lagrangian term responsible
for up quark masses.) Thus, one Higgs superfield has a spin-0 component
field that gives mass to down quarks, while the spin-0 component of
the other Higgs superfield yields up quark masses.\foot{In the common
nomenclature\refmark\hhg\ this means that the two-doublet model will
be type-II.}
Alternatively, one can also verify that two Higgs superfields are required
in order to complete the anomaly cancellations in the supersymmetric
context, where superpartners of the various gauge bosons are present.
Thus, there are five
physical Higgs bosons in the MSSM. They are: the $\hl$, the lightest
CP-even mass eigenstate;
\foot{The MSSM
Higgs potential is such that CP violation in the Higgs sector is not
possible.}
the $\hh$ the heavier of the two CP-even mass
eigenstates; the $\ha$, the single CP-odd state; and a charged Higgs pair,
$\hpm$. The resulting phenomenology is much richer than that of the MSM.

In the supersymmetric structure, the quartic couplings of the Higgs
fields become related to gauge couplings, and are no longer free
parameters. The quadratic mass terms are strongly constrained by
minimization conditions, and in the end only two parameters are required
to fully specify the Higgs potential.  These are normally taken to be
$\tanb$ (where $\tanb=v_2/v_1$ is the ratio of vacuum expectation values
for the two doublets) and $\mha$, the mass of the CP-odd scalar.
At tree-level, the masses and couplings of all the Higgs
bosons can be computed in terms of $\mha$ and $\tanb$. (In particular, the
mixing angle $\alpha$, arising in the diagonalization of
the neutral CP-even mass matrix, is determined.)  Additional
parameters are required to determine the one-loop corrections to the Higgs
masses, which can be large if $\mt$ and $\mstop$ (the stop squark mass)
are both large.  The basic results are well-known.
\Ref\hehperspectives{For a review and references, see H. Haber in
\perspectives, p. 79.}
The two most important points are:
\pointbegin
$\mhl\leq\mz+f(\mha,\tanb,\mt,\mstop,\ldots)$,
where $\ldots$ refers to generally less important parameters.
\point
For large $\mha$, $\mhh\sim\mhpm\sim\mha$, $\mhl$ approaches an upper
limit (which increases with increasing $\tanb$, $\mt$ and/or
$\mstop$), and the couplings of the $\hl$ become rather SM-like.
The approach of the $\hl$ couplings to SM-like values is, however, slow
enough that important and possibly measurable deviations will be present
even for $\mha$ values above several hundred GeV.

Constraints from existing data on the MSSM Higgs sector are few.
Data from LEP-I implies that $\mha\gsim 20\gev$ and $\mhl\gsim 40\gev$.
No constraint is currently placed on $\tanb$.  The $b \rta s\gamma$
decay branching ratio limit which is such a powerful constraint in
the non-SUSY two-doublet context is considerably weaker. In the strict
supersymmetric limit, $BR(b\rta s\gam)=0$.  Not surprisingly, there
is then a large region of SUSY parameter space such that there is no
inconsistency with current upper limits, even when $\mha$ is relatively
small (leading to $\mhp$ near its lower bound of $\sim\mw$)
\REF\bertolini{S. Bertolino, F. Borzumati, and A. Masiero, \npbj{294}
(1987) 321; S. Bertolini, F. Borzumati, A. Masiero, and G. Ridolfi,
\npbj{353} (1991) 591.}
\REF\bargiud{R. Barbieri and G.F. Giudice, preprint CERN-TH 6830/93
(1993).}
\REF\lopez{J. Lopez, D. Nanopoulos, and G. Park, preprint CTP-TAMU-16-93
(1993).}
\REF\oshimo{N. Oshimo, preprint IFM 12/92 (1992).}
\refmark{\bertolini-\oshimo}

In the last few years, it has been demonstrated that detection of at least
one of the MSSM Higgs bosons should be possible at the SSC, for almost all
choices of model parameters.
\Ref\susycontours{J.F. Gunion and L.H. Orr, \prdj{46} (1992) 2052.
Z. Kunszt and F. Zwirner, \npbj{385} (1992) 3.
H. Baer, M. Bisset, C. Kao and X. Tata,
\prdj{46} (1992) 1067.
V. Barger, K. Cheung, R.J.N. Phillips and A.L. Stange,
\prdj{46} (1992) 4914.}
This is often referred to as the `no-lose'
theorem for the MSSM Higgs sector.
\REF\gunperspectives{J.F. Gunion, in \perspectives, p. 179.}
\REF\ericeninetytwo{J.F. Gunion, preprint UCD-93-8, to appear
in Proceedings of the 23rd INFN Eloisatron Workshop {\it Properties
of SUSY Particles}, edited by L. Cifarelli and V. Khoze, World
Scientific Publishing.}
\REF\zwirnerreview{F. Zwirner, CERN preprint, CERN-TH-6792-93 (1993).}
For a review see, for example, Refs.~\gunperspectives,
\ericeninetytwo\ and \zwirnerreview.
To briefly summarize, we first note that
this statement relies only upon employing the
$gg\rta\h\rta ZZ^{(*)}\rta 4\ell$, $gg\rta\h\rta \gam\gam$ (and/or
$W\h+t\anti t\h\rta \ell\gam\gam X$) and $t\rta\hp b$ production/decay
modes (here $\h$ refers to $\hh$ or $\hl$). A region of parameter space that
is not covered by these modes arises if $\mt\sim 150\gev$ and $\mstop$ is
large (\eg\ $\mstop\sim 1\tev$); it comprises $110\lsim\mha\lsim160\gev$
and $\tanb\gsim 4$.\foot{If LEP-200 is considered as well, then the
region over which no MSSM Higgs boson is found at either LEP-200 or the SSC
is restricted to the same $\mha$ range, but $\tanb\gsim 10$.}
It should also be remarked that the $\ha$ is generally not observable at the
SSC in these modes (assuming $\tanb\gsim 1$), and that the $\hh$ can only be
detected in small regions of parameter space. For instance, $\hh\rta4\ell$
is detectable (at $\mt=150\gev$ and $\mstop=1\tev$) in a  tear-drop shaped
region roughly located  in the region $50\lsim \mha\lsim 2\mt$ and
$\tanb\lsim 5$. For $\mt\gsim 180\gev$ and $\mstop\sim 1\tev$, the region
over which $\hh\rta 4\ell$ is viable expands greatly, and the SSC alone
should be able to discover one of the MSSM Higgs bosons.

Clearly, our MSM discussion of the technical improvements, criticisms and
uncertainties for the $\gam\gam$ and $\ell\gam\gam$ modes will be highly
relevant to the MSSM Higgs discovery possibilities. However,
additional considerations enter in the MSSM extension, and these
will be reviewed.
The impact of the $gg\rta t\anti t \h \rta t \anti t b\anti b$
detection channel upon MSSM Higgs boson discovery will also be outlined.
In particular, expected $b$-tagging efficiency and purity
should allow us to detect $\hl\rta b\anti b$
in the above-noted window of parameter space where the $\gam\gam$ (or
$\ell\gam\gam$) and $4\ell$ modes are not adequate to detect one of the neutral
MSSM Higgs bosons.

It should be noted that the no-lose theorem relies primarily on the detection
of relatively light Higgs bosons.  Over much of parameter space,
only the $\hl$ is observed in the above-mentioned modes.  An implicit
assumption is that the $\hl$ does not decay to supersymmetric particle
final states.  While this is the most likely situation,
there do exist scenarios in which the lightest neutralino ($\cnone$)
is sufficiently light that $\hl\rta\cnone\cnone$ is kinematically
allowed, in which case this could be the dominant $\hl$ decay mode.
In such an instance, $\hl$ decays would be invisible. Thus, means for
detecting an invisibly decaying Higgs boson were developed, and will
be summarized.

But what about the heavier Higgs bosons of the MSSM? If $\mha\gsim 2\mz$,
the above detection modes may not allow us to observe any of
the other (approximately degenerate) Higgs bosons: the $\ha$, $\hh$
and $\hp$. Since the most characteristic signature for a two-doublet model
(in general) is the existence of a charged Higgs boson, the possibility
of $\hp$ detection in the $gg\rta \hp
b \anti t\rta t\anti t b\anti b$ final state was explored. For a significant
range of parameter space, it appears that
$b$-tagging efficiency and purity should
prove adequate to detect the $\hp$ in this mode.
Preliminary results will be summarized.

Of course, still more exotic models for the Higgs sector can be constructed.
Several popular extensions contain at least one neutral Higgs boson that
could have extremely weak coupling to fermion, but SM-like couplings
to gauge bosons.  The signals for such a Higgs boson and its associated
partners will be outlined.

More detailed reviews of the progress that has
been made prior to the Madison--Argonne workshops appears
in Ref.~\hhg, and in the various Snowmass SSC workshop proceedings and
LHC proceedings.  Where relevant, we will reference the appropriate material.
Here we will give  an overview of what has been accomplished during the
workshops, referring the reader for details to various individual
reports and recent preprints/publications.
In particular, we hope to establish the motivation for the projects
that were pursued, and put into proper context the progress made.

\vskip .15in
\noindent{\bf 2. The Standard Model}
\vskip .075in

As noted earlier,
substantial emphasis during these workshops was placed on improving our
ability to probe a relatively light $\hsm$.
Since it is not
at all unlikely that $\mhsm$ could fall in the range where the most
established detection modes rely on $\hsm\rta\gam\gam$ decays,
it is the $\gam\gam$ and $\ell\gam\gam$ modes upon which we first focus.

\FIG\widthratios{}
\topinsert
\vbox{\phantom{0}\vskip 5.0in
\phantom{0}
\vskip .5in
\hskip -10pt
\special{ insert user$1:[jfgucd.madison_argonne]widthratios_madisonargonne.ps}
\vskip -1.5in }
\centerline{\vbox{\hsize=12.4cm
\Tenpoint
\baselineskip=12pt
\noindent
Figure~\widthratios: The ratio of $\Gamma({\rm Higgs}\rta\gam\gam)$
computed for two different model choices for a number of cases.
In the case of the $\hsm$, the
ratio of the width predicted in the presence of an extra heavy generation
to that obtained in the MSM is shown.
For the $\hl$ and $\hh$, the ratio
$\Gamma(\h\rta\gam\gam)/\Gamma(\hsm\rta\gam\gam)$ as a function of
$\mh=\mhsm$ ($\h=\hl$ or $\hh$) is plotted.  For the $\hl$, $\mha$ is held
fixed
at $\mha=400\gev$; for the $\hh$, $\tanb$ is held fixed at $\tanb=7$.
Squark and slepton masses are determined
by choosing a common soft-SUSY breaking diagonal
mass, $\wtilde m$, with all off-diagonal mass terms set to zero.
Chargino masses are specified by the $M$ and $\mu$ parameters of
the mass matrix.  For both the $\hl$ and $\hh$, three cases are shown.
In the first \dots, $\wtilde m=300\gev$, $M=-\mu=300\gev$,
the lightest stop and the lightest chargino are both
always heavier than about $250\gev$.
In the second \dashes, $\wtilde m=300\gev$, $M=100\gev$, $\mu=-65\gev$,
the lightest chargino can be as light as about $43-53\gev$.
For the third case \dotdash, $\wtilde m=150\gev$, $M=100\gev$
$\mu=-65\gev$, the light stop eigenstate can be as light as $205\gev$.
The top quark mass is taken equal to 150 GeV for all calculations.
\REF\ghgamgam{J.F. Gunion and H.E. Haber, preprint UCD-92-22 (1992);
and \smv, p. 206.}
This figure is based on work from Ref.~\ghgamgam.}}
\endinsert

\FIG\ggxgamgam{}
\topinsert
\vbox{\phantom{0}\vskip 5.0in
\phantom{0}
\vskip .5in
\hskip -10pt
\special{ insert user$1:[jfgucd.madison_argonne]ggxgamgam.ps}
\vskip -1.5in }
\centerline{\vbox{\hsize=12.4cm
\Tenpoint
\baselineskip=12pt
\noindent
Figure~\ggxgamgam: The ratio of
$\Gamma({\rm Higgs}\rta g g)\times BR({\rm Higgs}\rta\gam\gam)$
computed for two different model choices for a number of cases.
In the case of the $\hsm$, we show the result in the presence of an extra heavy
generation divided by that obtained in the MSM.
For $\h=\hl$ or $\hh$, the ratio of the $\Gamma\times BR$ value
obtained in the MSSM divided by that computed for the $\hsm$ in the MSM for
$\mh=\mhsm$ is plotted.  All parameters, case choices and curve notations
are as for Fig.~\widthratios.
}}
\endinsert

\endpage
\bigskip
\noindent{\it 2.1. The $\gam\gam$ and $\ell\gam\gam$ Detection Modes}
\smallskip

The crucial $\hsm\gam\gam$ coupling derives from
the sum over all 1-loop diagrams
containing any charged particle whose mass arises from the Higgs
field vacuum expectation value, and is thus sensitive to many types of
new physics.  In particular, the 1-loop contribution
of a charged particle with mass $\gsim\mhsm/2$, approaches a constant value
that depends upon whether it is spin-0, spin-1/2, or spin-1. (The
contributions are in the ratio  $-1/3$ : $-4/3$ : 7, respectively.)
For a light Higgs boson, in the MSM the
dominant contribution is the $W$-loop diagram.  The next most important
contribution is that from the top quark loop, which tends to cancel part of
the $W$-loop contribution.  A fourth fermion generation with both a heavy
lepton, $L$, and a heavy $(U,D)$ quark doublet would lead to still further
cancellation. For $\mhsm\gsim 2\mw$, the $W$-loop contribution decreases,
and the heavy family ultimately dominates. To
illustrate, we show in Fig.~\widthratios\ the ratio of
$\Gamma(\hsm\rta\gam\gam)$ as computed in the MSM (with $\mt=150\gev$) to
that computed in the presence of an extra generation with $m_L=300\gev$ and
$m_U=m_D=500\gev$. For $\mhsm$ below $150\gev$, we see that the $\hsm\gam\gam$
coupling would be substantially below the MSM result.
However, at the same time the cross section for $gg\rta \hsm$
via heavy quark loops would be greatly enhanced. To first approximation,
the net event rate for the inclusive $\gam\gam$ mode remains
approximately unaltered, and we would still be able
to detect the $\hsm$ in the inclusive $\gam\gam$ mode.
\Ref\gunion{J.F. Gunion, work in progress.}
This is illustrated in Fig.~\ggxgamgam. There we compare the value obtained for
$\Gamma(\hsm\rta g g)\times BR(\hsm\rta\gam\gam)$
(which determines the number of inclusive
$\gam\gam$ events seen when the Higgs is produced via $gg$ fusion, followed
by decay to $\gam\gam$)
in the presence of an extra generation to that obtained in the MSM. We
see that over the $50\lsim\mhsm\lsim 130\gev$
range, where the inclusive $\gam\gam$ mode is potentially viable, the extra
generation result is between 1.1 and 1.5 times as large as the MSSM result.

On the other hand,
cross sections for the associated production
$W\hsm$ and $t\anti t\hsm$ processes,
responsible for the $\ell\gam\gam$ channel signal, would not be enhanced
due to the presence of an extra generation. The reduced $\hsm\rta\gam\gam$
coupling and branching ratio would then cause
the  Higgs signal in this channel to be substantially reduced,
and perhaps not even visible. Thus, the
observation of a Higgs signal in the inclusive $\gam\gam$ mode, in combination
with a reduced or absent signal in the $\ell\gam\gam$ mode, would imply
the presence of an otherwise unobservable heavy generation.

Of course, we should remark that the hypothesis of a light
$\hsm$ in the presence of a heavy fourth generation could easily be
inconsistent with vacuum stability in the context of the renormalization
group equations (see the discussion in the introduction) unless there
is new physics at a fairly low mass scale.

Because of the dominance of the $W$ loop contribution in the three family
case, the $\hsm\gam\gam$ coupling is also very sensitive to any deviations
of the $WW\gam$ and $WW\hsm$ couplings from SM values.
\REF\derujula{A. de Rujula \etal, \npbj{384} (1992) 3.}
\REF\perezt{M.A. Perez and J.J. Toscano, \plbj{289} (1992) 381.}
\refmark{\derujula,\perezt}\  The sensitivity to
anomalies in these couplings is potentially substantially greater than that
provided by LEP-I data, but only if the $\hsm$ can detected in the
$\gam\gam$ or $\ell\gam\gam$ modes and the
production cross sections reliably normalized.

Thus, the $\gam\gam$ and $\ell\gam\gam$
detection modes are of special importance. Discussion and contributions
to the workshop focused on two important aspects of these modes.
1) The impact of next-to-leading order (NLO) corrections.
2) The impact of experimental `reality'.

Let us discuss the latter first.  The implications of the CDF diphoton
measurements (performed at the Tevatron) for SSC backgrounds to
the $\gam\gam$ and $\ell\gam\gam$ modes were explored by R. Blair.
He considered two issues: a) to what extent can Monte Carlo estimates
of the backgrounds be relied upon; and b) using current
experimental data, what techniques can be
employed to better improve our estimates of such backgrounds,
especially backgrounds arising from jets that mimic an isolated photon.
His conclusions are summarized below.

With regard to a), there are several sources of uncertainty that will
remain right up to the turn
on of either the LHC or SSC.  First, our estimates of
the overall rate of the irreducible background from
prompt diphotons at LHC and SSC energies may not be
reliable.
Currently the rate measured at the Tevatron by CDF appears to exceed the
expected rate by more than a factor of two, even after
including NLO corrections.
\Ref\owensgamgam{J.~F.~Owens B.~Bailey and J.~Ohnemus, \prdj{46} (1992) 2018;
B.~Bailey and J.~F. Owens, \prdj{47} (1993) 2735.}\
This may indicate an even larger uncertainty for SSC/LHC experiments since the
region of $x$ of interest at CDF is roughly $2E_T/\sqrt{s}\approx .01$ and the
corresponding region for a 100 GeV $\hsm$ at the SSC would
be .005.  Thus, background estimates that use the
conventional distribution functions for computing
the QCD prompt diphoton rate may be more than a factor
of two low.  HERA measurements of the relevant quark and gluon distribution
functions should help to clarify this issue.
Another possible area where the background may be underestimated
is in the hadronic background.  At low $E_T$ the prompt photon cross section
as measured by CDF is also higher than that calculated from QCD.  Consequently,
if the background that comes
from a single photon plus a jet that fragments to a photon candidate is
computed based on QCD, it would be underestimated.

\FIG\rejection{}
\pageinsert
\vbox{\phantom{0}\vskip 8.25in
\phantom{0}
\vskip .5in
\hskip -80pt
\special{ insert user$1:[jfgucd.madison_argonne]blair_fig2.ps}
\vskip -1.85in }
\centerline{\vbox{\hsize=12.4cm
\Tenpoint
\baselineskip=12pt
\noindent
Figure~\rejection: Rate at which jets turn into photon candidates for
various values of isolation cut.  The candidates correspond to more
than one photon, presumably originating from hadronic decays.  Only
background candidates are included here; the prompt single photon
contribution has been subtracted off using a method similar to
that described in
\REF\abeetal{F. Abe \etal, \prlj{70} (1993) 1376.}
Ref.~\abeetal. Figure prepared by R. Blair.}}
\endinsert

While CDF data raises the uncertainties outlined above, it can also
be used to pin down other ingredients in the background calculations for
the SSC/LHC.  In particular,
using existing measurements (by CDF and D0) it is possible to determine how
often a jet will fragment to what appears to be an isolated photon candidate.
If this rate is coupled with the best estimates of jet-jet and photon plus jet
production at future colliders (including the uncertainties listed above) it
should be possible to quantify the corresponding level of background that will
be observed in a sample of diphoton candidates
for any given choice of isolation cut.
Additional detector dependent information can then be used to reject or
evaluate the non-prompt portion of the background.  This last step depends on
the intimate details of the detector and ultimately must
be evaluated once the detector is up and running, using in situ
sources of single photons or background (like $\gamma$'s from $\eta$ decays or
$\pi^0$'s) as CDF has.\refmark\abeetal\
The initial cut on isolation
typically reduces the hadronic background rates by more than two
orders of magnitude, as we shall shortly see, with isolation-only
rejection factors of order  $5-9\times 10^{-4}$ being typical.
CDF experience suggests that achieving an additional detector dependent
rejection factor of more than one order of magnitude will
not be easy. Since, adequate suppression of backgrounds
in the $\gam\gam$ ($\ell\gam\gam$) modes requires a total
rejection factor of $10^{-4}$ ($\sim 3\times 10^{-4}$)
for each jet, we see that the required rejection is fairly certain
in the $\ell\gam\gam$ case, but on the borderline for the inclusive
$\gam\gam$ mode.

Let us now return
to the rejection factor that can be achieved purely by an isolation cut.
At noted above, this can be directly measured by CDF.
The procedure is to determine
by other means (mainly shower profiles) the non-prompt component of
the event rate for a single isolated electromagnetic cluster
(\ie\ isolated photon candidate). By comparing
this rate to the inclusive jet rate (at the same $E_T$), the fraction
of jets that fake an isolated photon candidate can be computed
directly as the ratio of these two experimentally measured rates.
CDF results for the isolation cut rejection factor appear in Fig.~\rejection,
for a variety of $E_T$ values.
The isolation cut is defined by the fraction of additional energy observed
in a cone of size $\Delta R=0.7$ centered on the candidate.  As noted above,
quite decent isolation-only rejection factors of order $5-9\times 10^{-4}$
are achievable even for an additional energy fraction as large as 0.1.
The rejection factors
of Fig.~\rejection\ can be checked directly by CDF by using them
to determine the background to {\it two}-photon production coming
from one-photon plus jet production; simply multiply the latter by
the appropriate $E_T$-dependent rejection factor.
The resulting diphoton background can be compared
to that directly measured using shower profile information.
For a common choice of isolation-cone criteria, the agreement
is quite good.

Thus, the CDF results tend to provide a warning that backgrounds could
be larger than predicted using purely Monte Carlo computations,
even after NLO corrections are included. But, they
also indicate that the jet rejection factors required to eliminate non-prompt
backgrounds are achievable. In regard to the latter,
it should be noted that SSC/LHC analyses typically employ
an isolation criteria of $\lsim 5\gev$ of extra activity inside a cone
centered on the photon candidate. This corresponds to
an extra energy fraction of $\lsim 0.25$ for photons with $E_T$ of order
20 GeV, the lowest $E_T$ generally allowed in the SSC/LHC cuts.
This is a larger fraction than appears in Fig.~\rejection,
and typically employed in the above CDF analysis. Fig.~\rejection\
suggests that the jet-photon discrimination that can be achieved
on the basis of isolation alone, for
an extra energy fraction of $\lsim 0.25$, might be worse than $10^{-3}$.
An energy fraction of $\lsim 0.1$ would correspond
to rejecting extra energy that exceeds only
about 2 GeV. While this can be imposed at the Tevatron in the CDF
analysis of single photon and diphoton events, it might
be that the SSC events have more underlying minimum bias structure.  In this
case, the efficiency for such a cut on the Higgs signal might
not be very large. This is why the SSC/LHC analyses have adopted the more
conservative cut on extra hadronic energy in the isolation cone. Thus, the CDF
results for isolation rejection are not immediately applicable to the SSC/LHC
analyses. However, studies by the GEM collaboration
have suggested that realistic (in the SSC context) cuts yield
isolation and detector-dependent
rejection factors which combine to give a net jet rejection factor of order
$10^{-4}$.
\Ref\zhu{S. Mrenna, S. Shevchenko, X.R. Shi, H. Yamamoto, R.Y. Zhu,
preprint CALT-68-1856 (1993).}\ Similar claims have been made by
the LHC collaborations. Ultimately, this issue must be settled after
the SSC/LHC begins operation.

Let us now turn to the impact of NLO corrections.  In one contribution,
\Ref\baileygraudenz{B. Bailey and D. Graudenz, {\it Signal and Background
in NLO QCD for the Search of the Intermediate Mass Higgs Boson at the SSC},
preprint FSU-HEP-930801, these proceedings.}
B. Bailey and D. Graudenz have evaluated the NLO corrections to both
the inclusive $\gam\gam$ Higgs signal and the $\gam\gam$ continuum
background (including $gg\rta \gam\gam$ via the box diagram).  The computation
was done in the Monte Carlo context as delineated in Ref.~\owensgamgam,
for both signal and background.  This allows an accurate implementation
of the standard cuts (\eg\ $p_T\gsim 20\gev$, $|\eta|\lsim 2.5$
and hadronic energy inside isolation cone $\lsim 4-5\gev$, for the photons)
required to establish a signal. Their results
show that both signal and background are enhanced by K-factors of order
1.5 to 1.7, with the K-factor for the $\hsm$ signal being larger
than that for the background for $\mhsm\lsim 120\gev$.  Thus, not only
is the absolute event rate enhanced, but also the signal to background
ratio, $S/B$.  The result is that the number of standard deviations
of significance of the signal, $\nsd\equiv S/\sqrt B$, is larger according
to NLO computations than when only leading order (LO) computations are
employed, typically by a factor of order 1.3, rising to 1.4 for lower $\mhsm$
values near 80 GeV. This represents an important addition to our
confidence level for the viability of the inclusive $\gam\gam$ mode.

A second contribution, by D. Summers,
\Ref\summerscontribution{D. Summers, {\it The Search for a Light `Intermediate
Mass' Higgs Boson}, preprint MAD/PH/746, these proceedings.}
focuses on the associated $W\hsm$
and $t\anti t\hsm$ production modes, leading to the
$\ell\gam\gam$ signature for the Higgs boson. His reanalysis confirms earlier
results for signal and background rates for the two modes. In particular,
for an integrated luminosity of $L=10\fbi$ at the SSC, we recall that
the $W\hsm\rta \ell\gam\gam X$ mode has a rather low event rate,
and only by virtue of the $t\anti t\hsm$ production mode
is a viable signal achieved in the $\ell\gam\gam$ channel.
In comparison, at the LHC $L=100\fbi$ is required in order that the combined
signal be viable.  At the SSC,
for $L=100\fbi$ event rates in the $\ell\gam\gam$ channel coming from
$W\hsm$ and $t\anti t\hsm$ are both quite large (\eg\ 85 and 170 events,
respectively, for $\mt=140\gev$ and $\mhsm=110\gev$). Backgrounds
at $L=100\fbi$ from $W\gam\gam$ and $t\anti t\gam\gam$
(where $q\anti q$ collisions can be as important as $gg$ collisions)
are only 24 and 23 events, respectively (for $\Delta m\sim 0.03 \mhsm$).
Thus, there is an excellent chance that with high
luminosity at the SSC the $W\hsm$ and $t\anti t\hsm$ production
mechanisms can be separated (on the basis of much larger jet
activity for the $t\anti t\hsm$ mechanism), allowing a determination
of the ratio of the $\wp\wm$ and $t\anti t$ couplings of the $\hsm$.

Exactly how low in $\mhsm$ one can go in these modes is an important issue.
If it is only the irreducible backgrounds that are important, then,
with $L=10\fbi$ at the SSC, the $t\anti t\hsm$ mode remains viable
all the way down to the LEP upper limit of $\mhsm\sim 60\gev$.
For $L=100\fbi$ at the SSC, the $W\hsm$ mode also yields a viable signal
for such low masses.  This is nicely illustrated in the tables of
Ref.~\summerscontribution.  However, for $\mhsm$ below about $80\gev$,
the reducible backgrounds from $W\gam j+Wjj$ and $t\anti t \gam j+t\anti t jj$
production, where
the $j$'s appear to be photons, become much more problematical.
We will only know how low in mass these searches can be carried out
after the detector is up and running and the jet/photon discrimination
factor can be experimentally measured.

As Summers emphasizes, the above results are based on LO computations.
NLO corrections need to be examined.  It is well known that NLO corrections
to the $W\hsm$ process are closely related to those for inclusive on-shell $W$
production; one need only extend the calculation to an off-shell $W^*$
which decays to $W\hsm$.  Perturbative calculations to ${\cal O}(\alpha_s)$
\Ref\whnlo{R. Kleiss, Z. Kunszt and W.J. Stirling, \plbj{253} (1991) 269;
T. Han and S. Willenbrock, \plbj{273} (1991) 167; J. Ohnemus and W.J.
Stirling, preprint DTP/92/74; H. Baer, B. Bailey and J.F. Owens,
\prdj{47} (1993) 2730.} yield a $\sim10\%$ increase in the $W\hsm$ cross
section at NLO compared to LO.  At the workshops, P. Agrawal and C.P. Yuan
improved these calculations by taking into account the effect of multiple
soft gluon emissions on the kinematics of the decay products, using
the Collins-Soper-Sterman techniques for gluon resummation.
\Ref\css{J.C. Collins and D.E. Soper, \npbj{193} (1981) 381; {\it ibid.}
{\bf B197} (1982) 446; J.C. Collins, D.E. Soper and G. Sterman, {\it ibid.}
{\bf B250} (1985) 199.} They find that the net K-factor lies in the 5-15\%
range, depending upon the choice of
the parton distribution functions employed in the tree-level
calculation. (The tree-level result is larger for NLO distributions
than for LO distributions, leading to a smaller effective K-factor if NLO
distributions are used for both the tree-level and the full
${\cal O}(\alpha_s)$
computations.)  Also important is the extent to which the NLO
QCD corrections alter the distributions for the transverse momentum
and rapidity of the Higgs boson or of its photon decay products.
Agrawal and Yuan show that resummation effects do not affect the earlier
conclusions found in Ref.~\whnlo, namely that NLO corrections do not
alter these distributions relative to the naive LO computations.
\REF\aqy{P. Agrawal, J. Qiu, and C.P. Yuan, preprint MSUHEP-93/8.}
More details can be found in Ref.~\aqy.

NLO corrections to $t\anti t \hsm$
production are likely to be much larger than those for $W\hsm$,
with a K-factor of order 1.5 to 2
not being improbable.  However, an explicit computation has not yet appeared.
What about the background processes? There seems to be no reason to expect
the K-factor for the $t\anti t\gam\gam$ to be any larger than that for
the corresponding signal.  However, the situation for the $W\gam\gam$
background to $W\hsm$ is likely to be different.  In
\REF\ohnemus{J. Smith, D. Thomas, and W.L. van Neervern, {\it Z. Phys.}
{\bf C44} (1989) 267; J. Ohnemus, \prdj{47} (1993) 940.}
Ref.~\ohnemus\ it is shown that the NLO corrections to $q^\prime\anti q
\rta W\gam$ enhance the cross section by a factor of about 3.  This can
be understood by noting that the LO $W\gam$ cross section is suppressed
by the presence of radiation zeroes for certain angles of the final state
$W$ and $\gam$ with respect to the incoming quarks.  At next-to-leading order
these radiation zeroes are absent.  Thus the NLO cross section is of
standard size, while the LO cross section is artificially suppressed.
This same pattern is expected to continue in the case of the $W\gam\gam$
background to the $W\hsm\rta W\gam\gam$ signal.  Calculations of
the real radiation processes (\eg\ $qg\rta W\gam\gam q^\prime$)
yield cross sections of order 3.5 times the LO $q^\prime \anti q\rta W\gam\gam$
cross section even after a $p_T$ cut of $\geq 50$ GeV is imposed on the
extra jet.
\Ref\nlogamgam{H.Y. Zhou and Y.P. Kuang, \prdj{47} (1993) 3680.}\
Although the virtual diagram components of the NLO computation
have not yet been completed, it is clear that a net K-factor of order
2 to 3 is not out of the question.

While such an increase in the $W\gam\gam$
background would not be a particular problem for the overall $\ell\gam\gam$
signal (especially at the SSC where the $t\anti t\hsm\rta \ell\gam\gam X$
process is much
larger), one might worry that it would make the $W\hsm$ signal
much more difficult to isolate.  However, this need not be the case either.
To isolate the $W\hsm$ signal in the $\ell\gam\gam$ channel
we have already noted that one must veto
against extra jet activity in order to eliminate most of the $t\anti t \hsm$
component. In other words, what is crucial is the $W\gam\gam+0~{\rm jets}$
cross section.  As Summers notes in his contribution,
this will have far smaller NLO corrections
than fully inclusive $W\gam\gam$ production,
since the largest NLO corrections are due to real radiation processes.
A full assessment of the ability to isolate $W\hsm$ process requires the
computation of the virtual NLO contributions to $W\gam\gam$ as well as
a simulation of the $W\hsm$ signal events which includes minimum bias
low-$p_T$ jets and the gluon resummation effects discussed earlier.
Hopefully, a definition of `0 jets' can be found
which maintains high efficiency for the $W\hsm\rta W\gam\gam$ signal
events of interest, while reducing the $W\gam\gam$ background to more
or less the LO level.

In summary, the $\gam\gam$ and $\ell\gam\gam$ detection modes both
appear to be highly viable for the $\hsm$ in the MSM context,
provided appropriate detectors are available.  However,
as noted earlier,
if an extra heavy generation is present, the $\ell\gam\gam$
rates would be very suppressed and only the inclusive $\gam\gam$
channel is certain to be viable, and then only if 1\%
mass resolution and a jet-rejection factor of $\sim 10^{-4}$ are achieved.
As discussed, our ability to achieve the latter cannot
yet be regarded as proven.
Other scenarios could also impact these modes. For instance,
if the Higgs has only bosonic couplings (in which case a second Higgs
or some other mechanism must be responsible for giving mass to fermions)
then only the $W\hsm$ production mechanism would survive intact.
Since the $\hsm\rta\gam\gam$ branching ratio is likely to be quite large
(if not nearly 100\%) in such a scenario, the resulting $\ell\gam\gam$
event rate would be truly enormous and backgrounds would be irrelevant.
More discussion of this latter scenario will appear in a later section.

\bigskip
\noindent{\it 2.2. Detection of $t\anti t \hsm \rta t\anti t b\anti b$}
\smallskip

Although the importance of the $\gam\gam$ and $\ell\gam\gam$
modes is apparent, it would
be highly desirable to be able to detect a light $\hsm$ in its primary
$b\anti b$ decay mode.  Not only would this provide an alternative
detection mode, it would allow us to
normalize any Higgs signal that is seen in the $\gam\gam$ channels.
Normally, the primary decay mode, $\hsm\rta b\anti b$, is rejected
as having too large a QCD background.  In the MSM,
this is certainly the case for
inclusively produced $\hsm$'s.  (If an extra generation is present,
the $gg\rta \hsm\rta b\anti b$ rate might be sufficiently enhanced
that the mass of spectrum of two tagged $b$ jets would reveal
a signal in the inclusive case.)
However, at least two groups decided that
the associated production/decay mode $t\anti t\hsm\rta t\anti t b\anti b$
might hold promise if $\mt$ is large.  At this meeting we heard reports
from D. Wu
\Ref\dwu{T. Garavaglia, W. Kwong, and D. Wu,
{\it Search for Light Higgs via $t\anti t$ 2 Jets Channel},
these proceedings.}\
and J. Gunion
\Ref\dgvi{Based on
J. Dai, J.F. Gunion and R. Vega, preprint UCD-93-18 (1993), to appear
in \prlj{}.}\
which summarized the situation. Related work on the $W\hsm$ associated
production mode at the Tevatron was reported by S. Willenbrock.
\Ref\smw{A. Stange, W. Marciano, and S. Willenbrock,
preprint FERMILAB-PUB-93/142-T (1993),
and {\it Higgs Decay to Bottom Quarks at the Tevatron},
these proceedings.}

Let us focus first on the $t\anti t\hsm\rta t\anti t b\anti b$ process,
which is relevant for the SSC and LHC.
Two possible approaches to isolating the Higgs signal in the $t\anti t
b\anti b$ final state can be considered.  In the first, both top quarks
are required to decay leptonically and one looks at the two-jet mass
spectra for a Higgs bump. This was considered in both Ref.~\dwu\ and
Ref.~\dgvi.  In Ref.~\dwu\ the backgrounds considered were $t\anti t Z$,
with $Z\rta 2$~jets, and $t\anti t gg$, which is certainly the largest
background in the $t\anti t 2j$ channel. Approximate calculations
of signal and backgrounds were performed. Not
considered were the  $t\anti t q\anti q$ backgrounds.  The conclusion
of the analysis of Ref.~\dwu\ is that after appropriate cuts (especially
a cut requiring large $\ptr$ for any jet pair) a signal for $\mhsm\sim
100\gev$ could be detected. In Ref.~\dgvi\ the $t\anti t\hsm$ signal
and $t\anti t Z(\rta q\anti q)+t\anti t q\anti q$ backgrounds were computed
using exact matrix elements. The latter give a lower bound on the background
level --- the $t\anti t gg$ background can only worsen the situation.  In
order to compare to Ref.~\dwu, exactly the same cuts were employed.  The
result (for one SSC year of $L=10\fbi$) is that in the central $10\gev$ bin
centered on the $\mhsm=100\gev$
Higgs mass peak a signal of $S=221$ events is present over
a background of $B=2400$. (The latter figure includes combinatoric
backgrounds of all types, including those from the signal events
themselves.)  This yields a nominal statistical significance of $S/\sqrt
B=4.2$.  Even if an exact computation of the $t\anti t gg$ rate
were to only double $B=2400$ to $B=4800$, the signal becomes extremely
marginal.
In comparison to Ref.~\dwu, the signal rate of Ref.~\dgvi\ is
substantially smaller despite the inclusion of a `K' factor of 1.6
in the latter computations. The
estimate of Ref.~\dwu\ for the signal rate in a $10\gev$
bin centered at $100\gev$ is $S\sim 650$. The
$t\anti t gg$ background over this same mass
interval is estimated to be of order 9900 events.  This nominally yields
$S/\sqrt B\sim 6.5$, before inclusion of $t\anti t q \anti q$
backgrounds.

 \TABLE\ssclum{}
 \topinsert
 \titlestyle{\twelvepoint
 Table \ssclum: Number of $10\fbi$ years (signal event rate)
 at the SSC required for a  $5\sigma$ confidence level signal in four cases:
 I), II) --- 3 $b$ tagging with $(\ebtag,\emistag)=(30\%,1\%)$,
$(40\%,0.5\%)$; and
 III), IV) --- 4 $b$ tagging with $(\ebtag,\emistag)=(30\%,1\%)$,
$(40\%,0.5\%)$.}
 \bigskip

 \thicksize=0pt
 \hrule \vskip .04in \hrule
 \begintable
 Case | $\mt$\\$\mhsm$ | 80 & 100 & 120 & 140 \cr
 \ | 110 |   2.1( 331) &   3.5( 411) &   6.0( 458) &  29.1( 845) \nr
 I | 140 |   1.4( 324) &   3.0( 512) &   4.5( 486) &  22.4( 954) \nr
 \ | 180 |   0.3(  96) &   0.8( 175) &   2.9( 353) &  13.6( 747) \cr
 \ | 110 |   0.5( 191) &   1.0( 249) &   1.6( 275) &   7.6( 522) \nr
 II | 140 |   0.3( 170) &   0.6( 243) &   1.1( 266) &   5.3( 508) \nr
 \ | 180 |   0.1(  48) &   0.2(  97) &   0.6( 171) &   3.1( 375) \cr
 \ | 110 |  10.8( 135) &  15.9( 144) &  33.0( 186) & 170.0( 386) \nr
 III | 140 |   3.5(  95) &   5.0(  97) &  10.4( 122) &  55.0( 247) \nr
 \ | 180 |   1.0(  39) &   1.7(  45) &   4.2(  68) &  25.5( 160) \cr
 \ | 110 |   2.4(  94) &   4.1( 116) &   8.6( 153) &  46.0( 330) \nr
 IV | 140 |   0.6(  54) &   1.1(  66) &   2.6(  96) &  13.3( 189) \nr
 \ | 180 |   0.2(  23) &   0.4(  33) &   0.9(  46) &   5.5( 110) \endtable
 \hrule \vskip .04in \hrule
 \endinsert

The second approach to isolating the $t\anti t b\anti b$ final state is
to employ $b$-tagging.  At least three $b$'s must be tagged in order
to make headway against the general $t\anti t q\anti q+t\anti t gg$
backgrounds. This cuts down on the event rate to such an extent that
demanding double lepton tagging does not yield a viable signal.
In Ref.~\dgvi, a single lepton trigger is employed.  The other top quark
is allowed to decay either leptonically or hadronically.  The signal
and backgrounds were examined in the two cases where a) 3 or more
$b$'s are tagged and b) 4 or more $b$'s are tagged.  Crucial ingredients
are the efficiency for tagging a $b$ ($\ebtag$) and the probability that a
light quark or gluon jet is mis-identified (\ie\ mis-tagged) as a $b$
($\emisid$).  After appropriate kinematic cuts two cases were examined:
i) $\ebtag=0.3$ and $\emisid=0.01$, and ii) $\ebtag=0.4$ and
$\emisid=0.005$. After a few other cuts, none of which required
reconstruction of either the hadronically decaying $W$ or $t$ quark, the
invariant mass spectrum for two tagged $b$ jets was  examined for signal
vs. background. The results are encouraging for $\mt\gsim 130-140\gev$
and $\mhsm\lsim 110\gev$.  Table~\ssclum\ gives the number of
years required to see a 5 sigma signal (along with the associated number
of signal events in parentheses) for different Higgs masses and
top quark masses in the four scenarios: I=a)+i); II=a)+ii); III=b)+i);
and IV=b)+ii). Typically, the $\hsm$ is observable in $\lsim 3$
SSC years if $\mhsm\lsim
100\gev$ for $\mt=140$, and if $\mhsm\lsim 120\gev$ for $\mt=180\gev$.
(The actual peaks for $\mt=180\gev$ are quite dramatically visible
relative to the background --- the only issue is event rate.)

One virtue of this mode that should be apparent from Table~\ssclum\ is that an
$\mhsm$ value below $80\gev$ would actually yield an even more significant
signal. This is in contrast to the $\gam\gam$ modes which start to die
out below $\mhsm\sim 80\gev$ due to decreasing $\gam\gam$ branching
ratio and increasing $\gam\gam$ continuum backgrounds. As discussed earlier,
our ability to extend the $\gam\gam$ and $\ell\gam\gam$
mode searches to masses as low
as the $\sim 60\gev$ upper limit of LEP is crucially dependent upon
the level of reducible $\gam j$ and $jj$ type backgrounds.

What are the expectations for $b$-tagging?  The scenario of 30\%
tagging efficiency and 1\% mis-identification probability,
referred to as a) above, is quite close to that obtained
in the SDC TDR study of $b$ vertex tagging.
\Ref\sdctdr{Solenoidal Detector Collaboration Technical Design Report,
E.L. Berger \etal, Report SDC-92-201, SSCL-SR-1215, 1992, p 4.15-4.16.}
In fact, if the actual efficiencies computed in the SDC TDR are employed,
the results of Table~\ssclum\ improve somewhat.  Our only current
experience with $b$ tagging at a hadron collider is that
for the SVX vertex detector at CDF. These
$b$-tagging results were summarized by N.M. Shaw in his
contribution to these proceedings. Three different algorithms have
been explored so far, labelled `jet vertexing', `jet probability',
and `$d\phi$ clustering'.  All give similar results.
The net efficiency for tagging at least one of the two $b$
quarks in a $t\anti t$ event is about 22\% at $\mt=120\gev$, corresponding
to roughly 12\% per $b$-jet.
This efficiency includes SVX acceptance (which extends to roughly
$|\eta|\lsim 1$), but does not include any kinematic or high $p_T$
lepton cuts.  Given the limited $|\eta|$ range
and absence of momentum cuts, this number does not appear to be significantly
worse than what is obtained in the SDC TDR Monte Carlo study.
In any case, the SDC tracker is significantly better than that of CDF,
and so a better efficiency for the SDC detector should be anticipated.

Tagging of $b$-jets using the $\ell$ from the semi-leptonic
$b\rta c \ell\nu$ decay mode has also been employed by CDF.
For $\mt=140\gev$ a cut on the secondary lepton's $p_T$ of
2 GeV (relative to the jet axis) yields a net efficiency
(not including kinematic and primary high-$p_T$ trigger lepton cuts) for
tagging one or more $b$-jets of 23\%, \ie\ again of order 12\%
per $b$. (The fake tag rate is of order $10^{-2}$ per track.)
If this same efficiency were to apply in the SDC context, by combining
with a vertex tagging efficiency of order 30\% quite large
net $b$ tagging efficiencies at the SSC could be achieved, perhaps as
large as the 40\% assumed in scenario b) in the $t\anti t b\anti b$
channel study.

In any case, these respectable CDF $b$-tagging results and the
promising results outlined above for Higgs detection in
the $t\anti t b\anti b$ channel using $b$-tagging should
encourage the SSC/LHC detector groups to pay close attention to their
$b$-vertex
detectors. It is clear that maximizing the efficiency and purity of
$b$-tagging could have crucial benefits in the arena of Higgs detection.
Indeed, our group concluded that this was probably the single most
important focus for improvement of detectors that might still be possible
relatively late in the detector development and construction process.

The related work of Ref.~\smw, as reported by S. Willenbrock,
focuses on the $W\hsm+Z\hsm$ production modes, with $\hsm\rta b\anti b$,
at the Tevatron. (The current lower bound on $\mt$ rules out any
significant cross section for $t\anti t\hsm$ production at Tevatron
energies.) Leptonic decays of the $W$ or $Z$ (including $Z\rta \nu\anti \nu$)
are used to trigger the event (with appropriate cuts and isolation
criteria for charged leptons) and at least one of the $b$ quarks
is required to have $|\eta|<2$ and large enough $p_T$ that it could be
tagged.  Before including the actual efficiency for $b$-tagging,
the combined $W\hsm+Z\hsm$ cross section after all cuts and branching
ratios are included is about $0.2\pb$ at $\mhsm=60\gev$, falling to
about $0.07\pb$ by $\mhsm=100\gev$.  Two important backgrounds
are $Wb\anti b$ and $Zb\anti b$.  When integrated over an invariant
mass bin of $\pm 2 \Delta M_{b\anti b}$ (where $\Delta M_{b\anti b}\sim
0.8\sqrt{M_{b\anti b}}$) centered on $\mhsm$, these give a cross section
of very nearly the same size as the signal (after the same cuts \etc).
Backgrounds of similar size come from $ZZ$ and $WZ$ production.
Less troublesome are the backgrounds arising from $t\anti t$ production
by virtue of missing one of the $W$'s in the $t$ decays. $Wc\anti c$
and $Zc\anti c$ backgrounds could be comparable to their $b\anti b$ analogues,
depending upon the probability for the vertex tagger to mis-identify a $c$
quark as a $b$ quark. After all cuts \etc, the
$Wjj$ and $Zjj$ ($j=$ light quark or gluon) are about 100 times as
large as the $Wb\anti b$ and $Zb\anti b$ backgrounds. Thus, light-quark/gluon
vs. $b$ discrimination must be better than about $10^{-2}$.  Assuming
50\% efficiency  for detecting a displaced vertex from one or both of
the $b$ quarks in the signal (equivalent to 30\% efficiency per $b$ quark,
which is higher than the current efficiency
for the CDF SVX, but may be achieved by the time the main injector
is in operation), and a 1\% mis-tag probability for light
quark and gluon jets, one finds at $\mhsm=60\gev$ and $L=1000\pbi$
50/58 $W\hsm/Z\hsm$ events compared to $60/52$ $Wb\anti b/ Zb\anti b$
irreducible background events and $200/200$ $Wjj/Zjj$ mis-tagged events.
The resulting statistical significances in the two channels are of order
3 standard deviations. These signals will be further diluted by
inclusion of the $Wc\anti c/Zc\anti c$ backgrounds. Significant
signal enhancement would occur if the mis-tagging probability for
light quark and gluon jets could be reduced below 1\%.

\bigskip
\noindent{\it 2.3 A New Approach to the $\hsm\rta \lplm\nu\anti\nu$ Signal}
\smallskip

One of the modes suggested for detecting the $\hsm$ is
$pp\rta ZZ\rta\lplm\nu\anti\nu$, where the Higgs appears
as a resonance on a Jacobian background.  Unfortunately, there are QCD
background processes which mimic the final state that can only be removed
via stringent kinematic cuts.
In the process, a significant fraction of the signal is lost.
One approach to relaxing these stringent cuts and
extracting a better signal is that of tagging
one of the spectator jets that arise when the $\hsm$ is produced
via the $WW$ fusion mechanism.
\Ref\tagging{See, for example, V. Barger, K. Cheung, T. Han and D. Zeppenfeld,
preprint MAD/PH/757 (1993), and the contribution by K. Cheung to these
proceedings, and references therein.}
In a contribution to the workshops,
\Ref\duncanreno{M.J. Duncan and M.H. Reno, {\it Enhancing the Higgs
Signal in $pp\rta ZZ\rta\lplm\nu\anti\nu$}, these proceedings.}
Duncan and Reno have examined a method, that employs the distribution
of the final state charged leptons, which could enhance the gluon
fusion component of the signal in this same channel.
It relies on the fact that the $Z$'s coming from
the $\hsm$ are mainly longitudinally polarized, whereas all backgrounds
(\eg\ $qg\rta q Z$, where the $q$ disappears down the beam pipe,
as well as the irreducible $q\anti q\rta ZZ$ process)
yield transversely polarized $Z$'s. Also important is the fact that the $\hsm$
tends to be produced nearly at rest in $gg$ fusion.

If the center of mass of the $ZZ$ system could be determined,
then the distribution in $z=\cos\theta$ for the $\lplm$ pair
(where $\theta$ is the angle of one of the leptons with respect
to the boost direction of the parent $Z$)
could be determined. The shape
of this distribution would then yield a good estimate for the fraction
$f_L$ of events in which the visible $Z$ is longitudinally polarized.
However, since one of the $Z$'s decays invisibly, the $ZZ$ center of mass
cannot be determined.  Duncan and Reno define an alternative variable
$z^*\equiv 2|p_\ell\cdot\eps_L|/\mz$
(where $p_\ell$ is the momentum of one of the charged leptons
and $\eps_L=(|\vec p_Z|,\vec p_Z E_Z/|\vec p_Z|)/\mz$),
which coincides with $z$ when the $ZZ$ center of mass
and laboratory frames are the same. Since, in $gg$ fusion,
a heavy Higgs boson is
largely produced nearly at rest, $z^*$ should be a good approximation
to $z$ in the case of the signal contribution to this channel.
\def\avgz{\langle z^* \rangle}
If one computes the average value of $z^*$, then
theoretically $0.375<\avgz<0.5625$, the lower (upper) bound being reached
if only purely longitudinal (transverse) $Z$'s are produced.
Combining the $q \anti q\rta ZZ$ background and the $gg\rta\hsm\rta ZZ$
signal, a plot of $\avgz$ shows a pronounced dip as a function of
the transverse mass, $m_T$, in the vicinity of the Higgs resonance.
Of course, the vector boson fusion contributions to the $ZZ$ channel
(both resonance and continuum background) must be included, as
well as backgrounds such as that from $qg\rta q Z$; and errors must
be analyzed. This work is in progress.

\bigskip
\noindent{\it 2.4 Heavy Higgs and Strongly Interacting $W$ Scenarios}
\smallskip

As noted earlier, our ability to probe a strongly interacting EWSB
sector is reviewed in contributions by Bagger,
\Ref\bagger{J. Bagger, {\it Electroweak Symmetry Breaking at the
Supercollider}, these proceedings.}
Cheung,
\Ref\cheung{K. Cheung, {\it $W_LW_L$ Scattering at Hadronic Supercolliders:
Background Suppression}, these proceedings.}
and Morris.
\Ref\morris{D. A. Morris, {\it Prospects for Multiple Weak Gauge
Boson Production at Supercollider Energies}, these proceedings.}
Duncan Morris reported on work done in collaboration
with R.D. Peccei and R. Rosenfeld in which they considered the
phenomenology of a strongly interacting Higgs sector patterned after
hadronic pion physics.
\Ref\damorris{D.A. Morris, R.D. Peccei and R. Rosenfeld,
\prdj{47} (1993) 3839.} The central idea is to go beyond the
low energy theorems relating the behaviour of pions and
longitudinal weak gauge bosons and to assume, essentially by fiat, that
nonperturbative pion physics can serve as a model for a
strongly interacting $W_L$ ($W\equiv \wpm, Z$) sector.  By scaling up
multipion energy scales by a factor of $\sqrt{ v / f_\pi} \simeq 2600$
one can relate multipion production on the GeV scale to
multi-$W_L$ production on the TeV scale.
Though there is little reason to expect that a strongly interacting
Higgs sector should mimic pion physics in every detail, the assumed
equivalence permits a useful exploration of the phenomenological
consequences. A literal interpretation results in ${\cal O}(1-30)$~fb
cross sections at the SSC and LHC for the  production of multiple
$W_L$'s. When cuts are imposed to appropriately reduce
background processes such as multiple top quark production,
they find multi-$W_L$
signatures of multiple high-$p_T$ leptons and high-$p_T$ jets
with cross sections of ${\cal O}$(1~fb) at the SSC.
Multiple gauge boson production through generic perturbative
processes, strongly interacting Higgs sectors and a possible breakdown
in electroweak perturbation theory are also discussed in the contribution
by Morris.

Jon Bagger reviews the work of a large collaboration which explored
strong $W_LW_L$ interactions in a very wide class of models.
\Ref\baggeretal{J. Bagger, V. Barger, K. Cheung,
J. Gunion, T. Han, G. A. Ladinsky, R. Rosenfeld and C.--P. Yuan, preprint
FERMILAB-Pub-93/040-T (1993), to be published in \prdj{}.}\
More specifically, the processes $pp \rightarrow WWX$ were studied, and
the rates for the ``gold-plated'' channels, where
$W^\pm \rightarrow \ell^\pm \nu$ and $Z \rightarrow \ell^+
\ell^-$ $(\ell = e,\mu)$, were computed for each model.
Using a forward jet-tag, a central jet-veto and a back-to-back lepton cut to
suppress the Standard Model backgrounds, it was
demonstrated that the SSC and LHC have substantial
sensitivity to strong interactions in the
electroweak symmetry breaking sector.
Of course,
the channels examined, $\wp\wm\rta \lplm \nu\anti\nu$, $\wp Z\rta \lplm
\ell\nu$,
$ZZ\rta \lplm\lplm$, and $\wp\wp\rta \lplp\nu\nu$, do not {\it all} yield
adequate signals in 1-2 years of canonical SSC or LHC luminosity for
{\it all} models.
Instead, a significant signal can always be found
in the channels that most naturally complement the particular type
of model considered.  In particular, models with a resonance of definite
isospin are most easily probed using the $WW$ channels that have resonant
contributions from that same isospin. Non-resonant models are often
best probed in the $\wp\wp/\wm\wm$ channels.
Indeed, one important
conclusion is that different types of models can be distinguished
experimentally by determining the relative magnitude of the $LL$ signals in
the four channels listed above.
A large part of the work focused on the techniques required to suppress
reducible and, especially, irreducible backgrounds to a level such
that the low $LL$ signal event rates in the purely leptonic channels
can be isolated. In particular, the irreducible backgrounds
from production of $WW$ pairs with $TT$ and $LT$ polarizations
end up being most important, and the techniques developed in Ref.~\baggeretal\
are particularly
focused on suppressing them. Although the calculations do not include detector
effects, they should survive more sophisticated Monte Carlo
analyses.  In particular, the types of cuts employed should be
directly applicable in the experimental analyses
that will be performed when actual data becomes available.

In the contribution by K. Cheung, the types of cuts employed in
Ref.~\baggeretal\ are discussed in greater depth, emphasizing
the fact that in the $WW$ channel
the appropriate cuts for minimizing the irreducible
background from $TT+LT$ modes relative to the $LL$ signal are model
independent.  Detailed graphs explaining how the cuts were approximately
optimized are also given. As an aside, we also note
that there is now good agreement between the leading-log Monte Carlo
treatments and parton-level treatments
of $t\anti t$-related and other backgrounds. Thus, we have good
reason for confidence that the purely leptonic signals for a strongly
interacting $WW$ sector can be extracted.

Overall, it certainly appears to be possible to probe a strongly
interacting electroweak symmetry breaking sector at the SSC or LHC using
the ``gold--plated'' purely-leptonic modes.
Even if a light Higgs boson is found, it will be important
to measure the event rates at high $WW$ mass in all the various channels
in order to make certain that the Higgs boson
completely cures the bad high-energy behavior in all $WW$ scattering
subprocesses. The low event rates for the purely-leptonic final states imply
that of order 2-3 years of $10^4\pbi$ yearly luminosity
will be required to conclude that there is no obvious $W_LW_L$
enhancement in any of the four channels.
Because of the relative cleanliness of these final states, the option of
achieving this required integrated luminosity via enhanced instantaneous
luminosity should be strongly considered.

\bigskip
\noindent{\it 2.5 Conclusions for the Standard Model}
\smallskip

There seems to be little question that the SSC (and LHC) can probe
electroweak symmetry breaking for all the most attractive
scenarios that can be envisioned in the context of the Standard Model,
including the renormalization-group-motivated light intermediate-mass range,
as well as the case of a strongly interacting $WW$ sector.

\vskip .15in
\noindent{\bf 3. The Minimal Supersymmetric Model Higgs Sector}
\vskip .075in

As outlined in the Introduction, our goal is to establish a no-lose
theorem according to which we are guaranteed to see at least one
of the MSSM Higgs bosons regardless of the model parameter choices.
For much of parameter space, the $\hl$ is relatively light
and rather SM-like, and is the one that we could be certain of detecting.
Going beyond the no-lose theorem, it is desirable to be able to see one or more
of the heavier $\ha$, $\hh$ and $\hpm$ Higgs states.  However,
let us first focus on issues relevant to the no-lose theorem.
Of special concern will be the $\gam\gam$, $\ell\gam\gam$ channels
and the $t\anti t b\anti b$ channel.  Our discussion of these modes
will temporarily ignore the possibility of superparticle decays of a light
neutral $\hl$ or $\hh$.

\bigskip
\noindent{\it 3.1. The $\gam\gam$ and $\ell\gam\gam$ Detection Modes}
\smallskip

A particularly interesting question is the extent to which the
$\gam\gam$ widths of the MSSM Higgs bosons, especially that of the $\hl$,
depend upon
the SUSY context and/or superpartner masses. Some exploration of this
\REF\kileng{B. Kileng, preprint ISSN 0803-2696 (1993).}
issue has appeared in Refs.~\ghgamgam\ and \kileng.
Potentially, these widths are sensitive to  loops containing heavy charged
particles.  However, it must be recalled that supersymmetry decouples when
the SUSY scale is large. (In particular, superpartner masses come primarily
from soft SUSY-breaking terms in the Lagrangian and not from the Higgs
field vacuum expectation value(s).)  In Fig.~\widthratios\
several cases are illustrated.
In Fig.~\widthratios\ we show that if the MSSM parameters are
chosen such that all new particles beyond the SM are heavier than about 250 GeV
(technically we take $M=-\mu=300\gev$ for the charginos
and a common squark/slepton diagonal mass of 300 GeV), then the
deviation of $\Gamma(\hl\rta\gam\gam)$ from the corresponding SM value for
the $\hsm$ is less than 15\%. This is because of decoupling; as the SUSY
breaking scale and the scale of the heavier Higgs bosons become large,
the $WW$ coupling of the $\hl$ rapidly approaches its SM value
and the squark and
chargino loops become negligible. However, Fig.~\widthratios\
also shows that if the lightest chargino is
allowed to be as light as $\sim\mz/2$ (by taking $M=100\gev$ and
$\mu=-65\gev$),
so that it only just evades detection at LEP, then
the $\hl\rta\gam\gam$ width ratio can be suppressed to as
little as half the corresponding SM value.
In the case of the $\hh$, such a light chargino can greatly enhance the width.
This is also illustrated in Fig.~\widthratios, where $\Gamma(\hh\rta \gam\gam)/
\Gamma(\hsm\rta\gam\gam)$ is plotted in the same cases as considered
for the $\hl$.

The effect upon the $\gam\gam$ widths of the $\hl$ and $\hh$
of reducing further the common squark-slepton mass scale, while
maintaining a light chargino, is
also illustrated in Fig.~\widthratios. While modifications in the $\hl$ width
are not dramatic in the $\gsim 50\gev$ mass range of interest,
the $\hh\gam\gam$ coupling can be significantly suppressed when
squark and slepton masses are made light.

To determine the implications of these results
for the discovery potential for the MSSM Higgs bosons
in the $\gam\gam$ and $\ell\gam\gam$ channels requires additional ingredients.
Consider first the inclusive $\gam\gam$ mode.
In the case of inclusive $gg$ fusion production of the Higgs boson,
followed by $\gam\gam$ decay, the event rate is determined by
$\Gamma({\rm Higgs}\rta g g)\times BR({\rm Higgs}\rta\gam\gam)$.
In the MSSM the effects of including squark loops in the $gg\rta \h$
($\h=\hl$ or $\hh$) coupling computations are generally
numerically small. This was already evident from the early work of
\REF\ghii{J.F. Gunion and H.E. Haber, \npbj{278} (1986) 449.}
Ref.~\ghii. Potentially more important are modified couplings of the $\h$
to the fermions in the loops which dominate the $gg\rta\h$ coupling.
However, the $\hl t\anti t$ coupling tends to be
fairly SM-like for large $\mha$,
and, consequently, the $gg\rta \hl$ coupling tends to be rather SM-like.
Similarly, for $\mhh$
near its lower limit (the only region where the $\hh\rta\gam\gam$
mode is viable) the $\hh t\anti t$ and $gg\rta \hh$
couplings tend to be SM-like. It turns out that a more significant
impact on the inclusive $\gam\gam$ event rate occurs as a result of
modifications to the
$\h\rta\gam\gam$ branching ratios due to other channels.
For instance, the $\hl b\anti b$ coupling at $\mha=400\gev$
is still significantly above the SM value when $\tanb$ is large.  Thus,
$\Gamma(\hl\rta b\anti b)$ will be larger than in the SM and
$BR(\hl\rta \gam\gam)$ will be correspondingly suppressed.  In addition,
supersymmetric particle pair channels can enter into the $\h$ decays.
For the light chargino/neutralino sector scenario
considered in Fig.~\widthratios, this is possible even for the $\hl$
when $\mhl$ approaches its upper limit. Finally, in the case of the $\hh$,
if $\tanb$ is large then $b\anti b\rta \hh$ fusion production can be comparable
to $gg\rta\hh$ production.

To illustrate the importance of these additional effects, Fig.~\ggxgamgam\
shows
the ratio of $\Gamma(\h\rta g g)\times BR(\h\rta\gam\gam)$ (for $\h=\hl$ or
$\hh$) to the same quantity for $\h=\hsm$. The general viability of
the $\gam\gam$ mode for the $\hl$ (indicated by a ratio not too far below one)
is apparent, the exception being for $\mhl$ near its upper limit in the light
chargino/neutralino mode cases (when $\chitil\chitil$ modes become
kinematically allowed).  The very narrow region for $\hh$ discovery
in this mode is also apparent.  Clearly, the enhanced $\gam\gam$ width
found for the $\hh$ when charginos are light (see Fig.~\widthratios)
does not lead to any significant expansion of the very narrow region
for which $\hh$ discovery in the inclusive $\gam\gam$ mode is possible.

Turning to the $\ell\gam\gam$ channel, we first note that the $W\h$
and $t\anti t \h$ production rates are determined by the $\h\wp\wm$
and $\h t\anti t$ couplings, which in the relevant $\mhl$ and $\mhh$
mass ranges tend to to be fairly SM-like, as noted above.  The $\h\rta\gam\gam$
branching ratio is subject to the modifications outlined above.
Differences from the $\ell\gam\gam$ rate expected for the $\hsm$
are to be expected, although over much of the relevant mass range
such differences will be in the range of a factor of two to three.

Certainly, these results show that
the potential for MSSM Higgs discovery in the
$\gam\gam$ and $\ell\gam\gam$ channels
must be reassessed as LEP-II and, eventually,
the SSC provide information on the masses of the supersymmetric particles
that enter into the Higgs decays and the loop diagrams responsible for the
$\h\gam\gam$ and $\h gg$ couplings.
Even if supersymmetric decays of the $\hl$
and $\hh$ are not allowed, if either is detected in
these channels, the exact rate(s) could yield valuable self-consistency
tests for the masses and Higgs boson couplings of the supersymmetric particles.

The background to detection of a neutral MSSM Higgs boson in the inclusive
$\gam\gam$ mode are also affected by
superpartner particles.    In particular, there are squark loop contributions
to the $gg\rta\gam\gam$ background process
(which provides roughly 30-50\% of the continuum
background in the $\gam\gam$ inclusive channel). These were examined
by A. Djouadi and G. Belanger.  Their results are summarized below.

In the limit where the mass of the internal particle is much larger
than the invariant mass of the two photons $\hat s$, $\kappa\equiv
\hat s/(4 m^2)\ll 1$, the helicity amplitudes for loops of scalar
and fermion particles belonging to the fundamental triplet representation
are easily extracted. In terms of the angle $\theta$ of one of the
outgoing photons relative to an incoming gluon one finds:
\foot{6 boxes, 12 three-point and 3 two-point functions have to be
taken into account.}
\def\kp{\kappa}
$$\eqalign{
{\rm scalars} \ &: \ \half \sum_{\lambda_i} |M_{\lambda_1,\lambda_2,
\lambda_3,\lambda_4}|^2 = {272\over 2025} \kp^4 (3+\cos^2\theta)^2 \,,\cr
{\rm fermions} \ &: \ \half \sum_{\lambda_i} |M_{\lambda_1,\lambda_2,
\lambda_3,\lambda_4}|^2 = {4448\over 2025} \kp^4 (3+\cos^2\theta)^2 \,.\cr}
\eqn\djouadi$$
For the cross section one has to include the factors: 1/2 for two identical
photons in the final state; 2 for the color factor $(TrT^aT^b)^2$; color and
polarization averaging factors 1/8 and 1/2 for each gluon in the initial state.
The result is:
$$
d \sigma = {1\over 256} \alpha^2 \alpha_S^2 \sum _{\lambda_i} \sum_q
(e_q^2 |M_{\lambda_1,\lambda_2,\lambda_3,\lambda_4}|)^2 dPS\,.
\eqn\djouadii$$
Note that the final result for both the fermion and scalar amplitudes goes like
$1/m^4$. ($1/m^2$ should have been enough for the decoupling, but
the amplitude must be proportional to $F_{\mu \nu}^4$, \ie\
to $\hat{s}^2$,  and to get the correct dimension we need $1/m^4$ from the
loop.) Since the decoupling occurs so quickly, the approximation above
is likely to work even for invariant masses $\hat s$ that are just
slightly smaller than $4m^2$.
Even more noteworthy is the small size of a scalar loop
as compared to a fermion loop, which is roughly 16 times larger for the
same mass. Even if we set the squark masses all equal to the top quark
mass, the top contribution is still larger. Light quark loops make
contributions that behave as $\log (\hat s/m^2)$ (multiplied by some
$\pi^2$ factors), and will dominate by far both the top loop and the sum
of all the squark loops.  Thus, it seems that squark loop contributions
to $gg\rta \gam\gam$ are not numerically significant.

Of course, the $W\h$ and $t\anti t\h$ production
processes responsible for the $\ell\gam\gam$ detection modes are tree-level
processes, and will not be significantly influenced by superparticle loop
corrections.

We turn now to QCD corrections for the $gg\rta \h$ production
production processes.  As already emphasized in the SM $\hsm$ section,
when discussing the $\gam\gam$ inclusive mode, the NLO corrections
provide a substantial enhancement to the $gg$ fusion cross section,
and must be included.
For the $\hl$ and $\hh$ CP-even Higgs bosons, these are the same
as found for a $\hsm$ of the same mass (neglecting squark loops).
In a contribution to these proceedings
\Ref\ks{R.P. Kauffman and W. Schaffer, {\it Effective Lagrangian
for Gluons and Higgs Pseudoscalars}, preprint BNL-49448 (1993), these
proceedings; see also preprint BNL-49061.}
and another recent paper,
\Ref\djouadicor{M. Spira, A. Djouadi, D. Graudenz, and P.M. Zerwas,
preprint DESY-93-113 (1993).}
two groups have computed the NLO
corrections (in the large $\mt$ limit) for the CP-odd $\ha$.
They find that the real diagram corrections (\ie\ those arising
from diagrams in which a real gluon is radiated into the final state)
are the same for the CP-even and CP-odd Higgs bosons (when written
in terms of the respective LO cross sections).  The only difference
is in the virtual component of the NLO corrections.
For instance, in the contribution by Kauffman and Schaffer, the
virtual correction is characterized by the constant $N_c(\pi^2/3+2)$
for a CP-odd Higgs boson, as opposed to $(N_c\pi^2/3+5N_c/2-3C_F/2)$
for a CP-even Higgs boson.  These too are quite similar.  Thus, the
K-factor for CP-even and CP-odd Higgs bosons will be almost the same.

\FIG\surveyhbbi{}
\midinsert
\vbox{\phantom{0}\vskip 8in
\phantom{0}
\vskip .5in
\hskip -97pt
\special{ insert user$1:[jfgucd.rcsusyhiggs]hbb_susy_fig3.ps}
\vskip -2.15in }
{\rightskip=3pc
 \leftskip=3pc
 \Tenpoint\baselineskip=12pt
\noindent Figure~\surveyhbbi:
Discovery contours (at the $4\sigma$ level) in $\mha$--$\tanb$
parameter space for the SSC with $L=30\fbi$ and LEP-200 with
$L=500\pbi$ for the reactions: a) $\epem\rta
\hl Z$ at LEP-200;
e) $W\hl X\rta l\gam\gam X$;
g) $t\rta \hp b$; h) $t\anti t \hh$, with $\hh\rta b\anti b$;
and i) $t\anti t \hl$, with $\hl\rta b\anti b$.
Each contour is labelled by
the letter assigned to the reaction above,
on the side of the contour for which detection of the
particular reaction {\it is} possible.
The large $\times$ indicates the location of the window
where no MSSM Higgs could be discovered at LEP-II or the SSC
without processes h) and i).
We have taken $\mt=150\gev$, $\mstop=1\tev$ and neglected squark mixing.
Charginos and neutralinos are taken to be heavy.}

\endinsert

\bigskip
\noindent{\it 3.2. The $t\anti t b\anti b$ Detection Mode for $\hl$ and $\hh$}
\smallskip

As we have seen, the $t\anti t b\anti b$ channel, arising via associated
Higgs+$t\anti t$ production with Higgs$\rta b\anti b$
will be viable for a light SM $\hsm$.  Thus, it should be no surprise
that it is quite likely that the $\hl$ or $\hh$ of the MSSM
can also be detected in this way.
\Ref\dgvii{J. Dai, J.F. Gunion and R. Vega, preprint UCD-93-20 (1993),
to appear in \plbj{}.}
Indeed, over essentially all of parameter space either the $\hl$ or the $\hh$
has SM-like couplings. In addition, if $\mt$ is not too large,
the Higgs with SM-like couplings will be quite light, $\mh\lsim 120\gev$.
Thus, based on the analysis of the SM Higgs boson,\refmark\dgvi\
but correcting for branching ratio and production rate differences,
Ref.~\dgvii\ obtains the discovery contours in $\tanb-\mha$ parameter
space shown in Fig.~\surveyhbbi, in the case $\mt=150\gev$
and $\mstop=1\tev$. There, we have
adopted the same contour labels as used in J. Gunion and L. Orr,
Ref.~\susycontours, and Refs.~\gunperspectives\ and \ericeninetytwo,
in order to facilitate comparison with these previous treatments
of the no-lose theorem.
Scenario (A) refers to the notation established in Ref.~\ericeninetytwo;
it corresponds to the case in which decays of the Higgs bosons to
chargino and neutralino pair channels are not kinematically allowed.
In Fig.~\surveyhbbi, we have chosen to display only
those earlier-obtained contours that define the region (labelled by
the large $\times$) where the SSC would not be able to detect any of
the MSSM Higgs bosons using the $4\ell$, $\gam\gam$ (and/or $\ell\gam\gam$),
and $t\rta\hp b$ modes. (Discovery survey
plots in the above-noted references include a fuller selection of channels.)

{}From these contours the following conclusions can be drawn.
If $\mt\sim 150\gev$, then detection of the $\hl$
in the $t\anti t b\anti b$ mode will be possible for any $\mha\gsim 110\gev$,
the precise lower limit being slightly $\tanb$ dependent. (Note that
for such moderate to large values of $\mha$ it is the $\hl$ which is SM-like).
For $\mha\gsim 50\gev$ essentially all the way
up to the lower limit in $\mha$ at which $\hl$ detection becomes possible,
the $\hh$ is relatively light and is SM-like and can be detected in this mode.
\foot{There is a crossover in the vicinity of $\mha\sim 110\gev$
where the $\hl$ and $\hh$ interchange roles as being most SM-like.}
That is, either the $\hh$ or the $\hl$ can be detected in this way for
$\mha\gsim 50\gev$. This completely closes the no-lose theorem
parameter space hole (indicated by the big $\times$
in Fig.~\surveyhbbi) that arises if only $\gam\gam$, $\ell\gam\gam$
and $4\ell$ final states are employed. In
fact, by combining just the $t\rta \hp b$ detection mode and this $t\anti t
b\anti b$ final state mode, detection of at least one MSSM Higgs boson is
guaranteed to be possible at the SSC/LHC alone.

At $\mt\sim 200\gev$, the parameter space region for which the $t\anti
t+\hl,\hh\rta t\anti t+ b\anti b$ mode is viable is smaller; but this is
simply correlated with the fact that other decay modes, most notably the
$ZZ^{(*)}\rta 4\ell$ mode, of the $\hh$ and (rather heavy) $\hl$ acquire
larger branching ratios, and become viable over a large range of
parameter space. Thus, the no-lose theorem continues to hold, but a larger
set of final state modes most be employed. See Ref.~\dgvii\ for details.

Of course, it should be noted that this $t\anti t b\anti b$
mode is not viable for observing a heavy $\hh$ or heavy $\ha$;
at large $\tanb$, although the $b\anti b$ channel would dominate
the decays of the $\hh$ and $\ha$,
the $t\anti t \ha$ and $t\anti t \hh$ production
processes are suppressed. At small $\tanb$, the $\ha$ and $\hh$ decays
would be dominated by other modes such as $\hh\rta \hl\hl$
and $\ha\rta Z\hl$ for masses below $2\mt$ and by the $t\anti t$ channel
when kinematically allowed.

To reiterate, combining all modes, the SSC/LHC alone will detect
at least one and most probably several of the MSSM Higgs bosons.
The detected Higgs boson is most likely to be the $\hl$.
Indeed, it is not altogether unlikely that the $\hl$ can be detected in
all three of its most crucial decay channels, $ZZ^*$, $b\anti b$, and
$\gam\gam$, simultaneously. The other Higgs bosons are most likely to
be moderately heavy, in which case they could not
be detected using the modes discussed
so far.  We will return to these heavier Higgs bosons shortly.

\bigskip
\noindent{\it 3.3. Detection of an Invisibly Decaying Higgs Boson}
\smallskip

The above discussion does not take into account alternative
decay channels for the MSSM Higgs bosons.  Generally speaking,
dominance of the decays of a light Higgs boson
by invisible channels is quite possible in supersymmetric models.
In the case of the MSSM, decays to $\cnone\cnone$, where $\cnone$
is the lightest supersymmetric particle, can be dominant and would be
invisible if R-parity is conserved.
\Ref\susyinvisible{See, for instance: K. Greist and H.E. Haber,
\prdj{37} (1988) 719; A. Djouadi, J. Kalinowski,
and P.M. Zerwas, \zpcj{57} (1993) 569; and J.L. Lopez, D.V. Nanopoulos,
H. Pois, X. Wang, and A. Zichichi, preprint CERN-PPE-93-16 (1993).}
In the MSSM, $\cnone\cnone$
dominance is possible for both the $\hl$ and the $\ha$.
In supersymmetric models with spontaneously
broken R-parity, the dominant decay mode of the lightest scalar
Higgs boson is predicted to be $\hn\rta JJ$, where $J$ is the
(massless) Majoron.%
\Ref\majoron{J.C. Romao, F. de Campos, and J.W.F. Valle,
\plbj{292} (1992) 329.}\
$J$ interacts too weakly to be observed in the detector.
In such models, the decays of the second lightest scalar Higgs boson
can also be predominantly invisible, the two most important modes being
$JJ$ and $\hn\hn(\rta JJJJ)$.
\Ref\majoronii{J.C. Romao, J.L. Diaz-Cruz, F. de Campos, and J.W.F. Valle,
preprint FTUV-92-39 (1992).}

Two possible detection modes for an invisibly decaying Higgs boson
can be envisioned at a hadron collider.  Associated $Z\hn$ production
was considered in
\REF\fkane{S. Frederiksen, N. Johnson, G. Kane, and J. Reid,
SSCL-preprint-577 (1992), unpublished.}
Ref.~\fkane, with the rough conclusion that a viable
signal for $\hn\rta I$ ($I$ being any invisible channel) can be detected
for $\mhn\lsim 150\gev$ provided the $\hn ZZ$ coupling is Standard Model
(SM) strength and $BR(\hn\rta I)\sim 1$. As part of the workshops,
an alternative based on associated production of
top plus anti-top plus Higgs was considered.
\Ref\guninvisible{J.F. Gunion, U.C. Davis preprint, UCD-93-28 (1993),
submitted to \prlj{}.} There, it is found
that if the top quark is not too light
($\mt\gsim 130\gev$) and if $BR(\hn\rta I)\sim 1$, then a viable signal for
$\hn\rta I$ can be extracted for $\mhn\lsim 250\gev$ when the $\hn t\anti t$
coupling is of SM strength. Clearly, the $Z\hn$ and $\tth$ modes are
complementary in the sense that they rely on the vector boson
vs. fermion couplings, respectively, of the Higgs boson.
For a CP-even Higgs boson,
which has both types of coupling, both modes tend to be viable,
but for a CP-odd Higgs boson the $ZZ$ coupling is
absent at tree-level and only the $\tth$ mode outlined below
could lead to a visible signal.

The procedure for the $t\anti t \h$ case is quite simple.
One triggers on $t\anti t \hn$ events via an isolated lepton
($e$ or $\mu$) from one $t$ decay.
In order to further single out events containing a $t\anti t$ pair,
at least one of the $b$-quarks must be vertex tagged.
The procedure is that outlined earlier in the $t\anti t b\anti b$ channel case.
Results given below are for $\ebtag=0.3$.
Mis-identification backgrounds are not significant so long as
$\emistag\sim0.01$, and the corresponding
number for $c$-quark jets is of order 0.05.

The invariant mass of each pair of
jets, $\mtwo$, is computed and at least one pair {\it not containing the
tagged $b$-quark} is required to have
$\mw-\delmw/2 \leq\mtwo\leq\mw+\delmw/2$. In addition,
each pair of jets satisfying this criteria is combined
with the tagged $b$ jet(s) to compute the three-jet invariant mass, $\mthree$.
One demands that $\mt-\delmt/2\leq\mthree\leq\mt+\delmt/2$ for at least
one $bjj$ combination.  Together, these two cuts greatly reduce
the likelihood that the second top in a $t\anti t$ event
can decay leptonically and satisfy all our criteria. In fact,
if both $t$'s decay leptonically,
to leading order only $t\anti t g$ events in which the non-tagged
$b$-quark and the $g$ combine to yield an invariant mass
near $\mw$ can pass the $\mtwo$ cut, and this non-tagged $b$ plus
the $g$ jet must combine with the tagged $b$ to give a mass near $\mt$.
If mass cuts of $\delmw=15\gev$ and
$\delmt=25\gev$ are used, only a small fraction of
signal events are eliminated when typical
SDC jet and lepton energy resolutions are employed,
whereas the reducible backgrounds are significantly decreased.

Finally, to reveal the invisibly decaying Higgs, one employs the
missing transverse momentum, $\ptmiss$, for the event and computes
$\mmissl$, the transverse mass obtained by combining the
transverse components of the missing momentum and the lepton momentum,
$\mmissl^2\equiv(\etmiss+\etell)^2-(\ptmiss+\ptell)^2$.
The $t\anti t\hn$ events events of interest
are characterized by very broad distributions in $\mmissl$ and $\etmiss$.
Cuts on both variables are made.

There are several sources of background.  The most obvious is the
irreducible background from $t\anti t Z$ events in which $Z\rta \nu\anti\nu$.
This background will be denoted by $\ttz$; its $\etmiss$ and $\mmissl$
distributions are very much like those of the signal.
The important reducible backgrounds all derive from various tails related
to $t\anti t$ or $t\anti t g$ events.   It is useful to artificially separate
the $t\anti t (g)$ backgrounds into two components. The first, and most
important, component is that already alluded to above,
where both top quarks decay leptonically. As already
noted, only $t\anti t g$ events can possibly satisfy the additional
cuts imposed.  We shall refer to this background as the $\ttglnu$
background. The background from $t\anti t$ plus
$t\anti t g$ events, in which only one
$W$ coming from the $t$ quarks decays leptonically, can be significant
because of semi-leptonic decays of the $b$ quarks present, and
will be denoted by $\ttbdecay$.

The strategies required to control the $\ttglnu$ and $\ttbdecay$ backgrounds
are more or less `orthogonal'.
The $\ttglnu$ background is easily reduced to a level below that
of the $\ttz$ background by a cut on $\etmiss$.
And, it is easy to find an $\mmissl$ cut which reduces the
$\ttbdecay$ background to a negligible level while retaining
much of the $\tth$ signal (and remaining $\ttz$ background).

 \TABLE\sscluminv{}
 \topinsert
 \titlestyle{\twelvepoint
 Table \ssclum: Number of $10\fbi$ years (signal event rate)
 at the SSC required for a $5\sigma$ confidence level signal,
for $\mt=140\gev$, $\etmiss>200\gev$ and
$\mmissl>150\gev$ if $\delmw=15\gev$ and $\delmt=25\gev$.
$BR(\hn\rta I)=1$ is assumed.}
 \bigskip
 \hrule \vskip .04in \hrule
 \thicksize=0pt
 \begintable
  $\mt$ \\ $\mhn$ (GeV) | 60 & 100 & 140 & 200 & 300 \cr
  110 |   1.8( 26) &   2.6( 30) & 3.3( 34) &
  8.4( 55) & 15.5( 74) \cr
  140 |   0.3( 19) &   0.4( 22) & 0.7( 29) &
  1.4( 41) & 2.9( 60) \cr
  180 |   0.2( 27) &   0.3( 34) & 0.6( 45) &
  1.3( 68) & 3.9( 118) \endtable
 \hrule \vskip .04in \hrule
 \endinsert

The number of SSC years required for a $\nsd=5$
sigma significance of the signal compared to background for $BR(\hn\rta I)=1$
is given in Table~\sscluminv, for a variety of $\mhn$ and $\mt$ values.
The statistical significance $\nsd$ is computed as $S/\sqrt B$.
For any given integrated luminosity, $S$ is the total $\tth$ event rate
and $B$ the total $\ttz+\ttglnu+\ttbdecay$
event rate, with $\etmiss>200\gev$ and $\mmissl>150\gev$
(and all other cuts) imposed.
Also given (in parentheses) is the associated number of signal events
($S$).  The associated number of background events ($B$)
can be obtained from the relation $B=S^2/25$.

{}From this table, it is immediately apparent that
detection of an invisibly decaying Higgs boson should be possible
within 1 to 2 SSC years for $\mhn\lsim 200-250\gev$
if (as assumed in these calculations)
its coupling to $t\anti t$ is of Standard Model strength and
$\mt\gsim 130\gev$.
(The required number of years for non-SM coupling is obtained
simply by dividing the results of Table~\ssclum\ by the ratio of the
$t\anti t$ coupling strength squared to the SM strength squared.)
For $\mhn\gsim 300\gev$, the $\tth$ event rate drops to a lower level
such that more than 2 SSC years are required. However, it is rather
unlikely that invisible decays would be dominant for a Higgs boson
with mass above 200 GeV or so.  The table shows that
detection of the $\hn$ generally becomes easier
for heavier top quark masses. This is because the $\ttz$ and
$t\anti t (g)$-related
backgrounds are smaller and the signal rates somewhat larger
than for smaller $\mt$.  However, even for $\mt$ as low as $110\gev$,
Higgs bosons with mass below about 140 GeV that decay invisibly
should be detectable in less than 3 SSC years.

Of course, the $\nsd$ values quoted assume that the normalization
of the expected background will be well-determined by the time
that the experiments are performed.  This will require a good
understanding of the missing energy tails as they actually appear
in the detectors, calculation of the higher-order QCD
corrections that were only estimated,
and accurate knowledge of the parton (especially gluon)
distribution functions.  With the availability of HERA data, and through
the analysis and study of $t\anti t$ events in the actual detectors,
it is likely that uncertainties in the relevant backgrounds can be brought down
to the 20\% level by the time that adequate luminosity has been
accumulated that an invisible Higgs signal would become apparent at the SSC.

The above study was performed for a $\hn$ that is a CP-even
Higgs mass eigenstate.
The $\tth$ rates would be somewhat different as a function of
$\mhn$ for a mixed CP or CP-odd eigenstate. However, we do not
anticipate that the results for such cases would differ by very
much from those obtained here.

The immediate relevance of these results to the MSSM is somewhat model
dependent.  In ongoing work, H. Pois and J. Gunion have begun exploration
of a variety of grand unification scenarios, in particular the
so-called no-scale scenario.  Especially in this latter, it is quite
possible for neutralinos to be light enough
that SUSY decay modes of the $\hl$ are allowed. Typically,
the $b\anti b$ and $\cnone\cnone$ modes then compete with one another.
After also accounting for the $t\anti t \hl$ coupling,
it is found that one can
guarantee detection of the $\hl$ at the 5 sigma level
after 3 SSC years in either the $t\anti t b\anti b$
channel discussed earlier or the invisible decay channel, if not both.

\bigskip
\noindent{\it 3.4. Other Superparticle Decay Modes}
\smallskip

More generally, when allowed, $\chitil\chitil$
(where $\chitil$ represents a chargino or neutralino)
decay modes of the MSSM Higgs are substantial, and often dominant.
Some work on this subject at the SSC/LHC has appeared in
\REF\baersusy{H. Baer, M. Bisset, D. Dicus, C. Kao, and X. Tata, \prdj{47}
(1993) 1062.}
Refs.~\baersusy, \gunperspectives, and \ericeninetytwo.
In particular, the above references show that a significant
branching ratio for $\chitil\chitil$ decays would make detection
of the $\hh$ in the $4\ell$ mode impossible.  At the same time, however,
these same decays can lead to final
states containing multiple leptons, which would be relatively free of
background. The preliminary exploration of Ref.~\baersusy\ indicates
that detection of the excess events in such channels due to
Higgs$\rta\chitil\chitil$ decays may be possible.

\bigskip
\noindent{\it 3.5. Detection of the $\hp$ in the $t\anti t b\anti b$
Final State}
\smallskip

Of course, if $\chitil\chitil$ channels do not dominate the decays
of the $\hh$, $\ha$ and $\hp$, detection of these latter Higgs bosons
must be via SM particle channels.  As emphasized earlier, this
appears to be a non-trivial task.
One mode that has been investigated is $\hh,\ha\rta \tauptaum$,
originally suggested by Kunszt and Zwirner,
\Ref\kzfirst{Z. Kunszt and F. Zwirner, in \aachen, Vol.II, p. 578.}
It potentially becomes viable at large $\tanb$, where
the $gg\rta b\anti b+\ha,\hh$
production rates are greatly enhanced and $b\anti b$ decays have $\sim
90\%$ branching ratio (in the absence of $\chitil\chitil$ decays).
In the L3P simulation study,
\Ref\rubbia{See for example the review by A. Rubbia, \hawaii.}
$\ha$ and $\hh$ detection in this mode is claimed to be
viable for all $\mha\gsim100\gev$ and $\tanb\gsim 7$.

What about the $\hp$?  In many ways a charged Higgs boson is the hallmark
of a truly non-minimal Higgs sector, and in particular of two-doublet models
such as the MSSM. In contrast, the presence of
more than one neutral Higgs boson could
be due to additional singlet Higgs representations beyond the single
MSM doublet.  Work was begun during the course of these workshops,
\Ref\gunhp{J.F. Gunion, work in progress.} to assess the possibility
of detecting a charged Higgs boson in the $gg\rta b\anti t \hp\rta
b\anti t t \anti b$ production/decay mode.  A preliminary report
on this work is given here. Two-doublet model-II type fermion
couplings\refmark\hhg\ for the charged Higgs are assumed.

\FIG\hplus{}
\topinsert
\vbox{\phantom{0}\vskip 5.0in
\phantom{0}
\vskip .5in
\hskip -10pt
\special{ insert user$1:[jfgucd.madison_argonne]hplus_argonne.ps}
\vskip -1.45in }
\centerline{\vbox{\hsize=12.4cm
\Tenpoint
\baselineskip=12pt
\noindent
Figure~\hplus: $dN/d\mthree$ is plotted as a function of $\mthree$
for: the $gg\rta b\anti t \hp+\anti b t \hm$ signal (solid);
the $gg\rta t\anti t b\anti b$
background (dots); and the $t\anti t g$ mis-tagged background (dashes).
For this plot we have taken $\tanb=1$ and $\mt=140 GeV$ and integrated
luminosity of $L=10\fbi$ at the SSC.
Signal curves are given for $\mhp=180$, $200$, $300$, $400$ and $500\gev$.
Results do not include any QCD K-factors for the $t b \hpm$ signal
or $t\anti t b\anti b$ background.  No additional K-factor
for the $t\anti t g$ background is appropriate.
}}
\endinsert

Events are tagged by requiring that one of the $t$'s decay to a leptonically
decaying $W$.  Three $b$-jets are required to be tagged. Cuts and efficiencies
for these tags
are the same as those used in previous studies of this final state.
The second $W$ from $t$ decay is required to decay hadronically. $\mtwo$
and $\mthree$ cuts are imposed as described in Section 3.3.
Finally, a plot of the $\mfour$ mass distribution is made, where both
$b$'s are required to be tagged and the two $j$'s must not have been
tagged. The only important backgrounds, after such cuts, turn out
to be the $t\anti t b\anti b$ continuum QCD background, and $t\anti t g$
where the $g$ is mis-tagged (with 1\% probability) as a $b$-jet.
The $t\anti t Z$, with $Z\rta b\anti b$, background is much smaller than
either. Typical results for the $\mfour$ distribution are shown in
Fig.~\hplus.  In this figure, the $gg\rta b\anti t \hp+\anti b t \hm$ signal
and $gg\rta t\anti t b\anti b$ backgrounds are computed at LO.  For
three $b$ tagging, there will be a large K-factor by
which these should be multiplied. By comparing the exact computation
of the $2\rta 2$ processes
$gb\rta t\hm+g\anti b\rta \anti t \hp$ as performed in
\REF\everyone{J.F. Gunion, H.E. Haber, F.E. Paige, Wu-Ki Tung, and
S.S.D. Willenbrock, \npbj{294} (1984) 621.}
Ref.~\everyone\ to the results obtained from the LO
$gg\rta b\anti t \hp+\anti b t \hm$ $2\rta3$ calculation, before any cuts,
this K-factor is estimated to be in the range 2-2.5. We have employed
the $2\rta 3$ computation in order to more correctly account for $b$-tagging,
including multiple tag possibilities and kinematic cuts.

Fig.~\hplus\ shows that a significant $\hp+\hm$ signal will be present
(especially if it is appropriate to multiply the $tb\hpm$ signal
(and $t\anti t b\anti b$ background) by a K-factor of order 2.5).
At $\mt=180 \gev$, the same type of plot would show
$\hpm$ signals that are dramatically above background at $\tanb=1$.
For $\mt=110 \gev$, $\hpm$ detection in this mode would be very difficult.
The magnitude of the signal for
an $\hpm$ of any given mass depends significantly on
$\tanb$. Assuming model-II couplings,
as $\tanb$ increases above 1 the signals decrease to a minimum
level at $\tanb\sim 5-6$ rising back up to the $\tanb=1$ level
for $\tanb\gsim 20$. For $\tanb$ values above this level, it should be
possible to detect the $\hp$ in this manner
if $\mt\gsim 140\gev$.  Since
in many GUT scenarios the large value of $\mt$ is correlated with
a large $\tanb$ value, this mode holds considerable promise.
More details on precise statistical significances \etc\
will appear in a forthcoming paper.\refmark\gunhp\
Of course, it should be noted that the above discussion is largely
independent of whether or not the $\hp$ is part of the MSSM
Higgs sector, or simply the charged member of a more general
Higgs doublet with couplings to fermions of model-II type.

\bigskip
\noindent{\it 3.6. Conclusions for the MSSM}
\smallskip

We have made some remarkable strides in MSSM Higgs detection in the
last year.  With the addition of the $t\anti t b\anti b$
and $t\anti t+$invisible detection techniques,
the no-lose theorem has no remaining loop-holes. Detection of one
of the MSSM Higgs bosons (most likely the $\hl$) will be possible at
the SSC for all possible MSSM parameter choices.
Further, progress on detection of the heavier Higgs bosons of the
MSSM continues.

\smallskip
\noindent{\bf 4. A Higgs Boson with Negligible Fermionic Couplings}
\smallskip

It is not impossible that the symmetry breaking mechanism responsible
for giving masses to the gauge bosons is separate from that which
generates the fermion masses.  In this case, there could exist
a neutral Higgs boson, $\h$, which couples at tree-level only to gauge bosons
and not to fermions.  Such a neutral Higgs boson can also arise
in the context of triplet Higgs representations.\refmark\hhg\
There, the triplet Higgs bosons do not have the appropriate quantum numbers
to allow tree-level fermion couplings, but they can couple to the gauge
bosons.  In both cases, fermion couplings will be generated by one-loop
diagrams. Fine-tuning is required to enforce small fermion couplings to all
orders in perturbation theory. Nonetheless, it is amusing to consider
the phenomenology of a neutral $\h$ with only tiny $f\anti f$ couplings.
Also important is the phenomenology forced upon other members of the Higgs
sector by choosing parameters such that the $\h$ be of this special character.

Turning first to the phenomenology of the $\h$, we note that,
if such an $\h$ has large enough mass, its decays to $WW^{(*)}$ channels
will certainly be dominant, and the phenomenology for the $\h$
would not be very different from that of the $\hsm$ (at the same mass).
But, far enough below $WW^*$ threshold the $\gam\gam$ decay of
the $\h$ (as induced by the $W$ loop diagram) will be dominant,
rather than $f\anti f$ channels. The exact value of $\mh$ below which
$\gam\gam$ decays of the $\h$ begin to dominate
is model dependent, but is generally below $100\gev$.
Since the $ZZ$ coupling of the $\h$ is by assumption not suppressed,
if the $\h$ were light enough it could
be produced  with significant rate in the $Z\rta Z^*\h$
channel at LEP and there would be
an excess of $\nu\anti\nu\gam\gam$, $\lplm\gam\gam$ and $q\anti q\gam\gam$
events there. If we assume that there is no such excess, then
most likely $\mh\gsim 60\gev$.

We have already noted that the signature for an $\h$ with mass in
the $60-130\gev$ range that decays primarily to $\gam\gam$ would
be a truly spectacular enhancement in the $W\h\rta \ell\gam\gam X$
channel.  There would be no difficulty in seeing such an $\h$
at the SSC. In addition, there would be a not-insignificant
inclusive $\gam\gam$ signal coming from $WW$ fusion production of
the $\h$ followed by $\h\rta\gam\gam$ decay.  Even though
the $WW$ fusion cross section is
normally neglected in comparison to $gg$ fusion
when discussing a light $\hsm$, the $WW$ fusion production
rate is only about a factor of 10 lower.  If the $\gam\gam$ branching ratio
of the $\h$ is enhanced by a factor of 10 to 1000
compared to the SM $\hsm$, the signal in the
$qq\rta qq \wp\wm\rta qq \gam\gam$ channel would be easily
visible. Ref.~\smw\ has examined the feasibility of detecting
the $\h$ in the $W\h$ mode even at the Tevatron.
Their conclusion is that with $L=100\pbi$, the $W\gam\gam$ channel
can be used to detect such an $\h$ for $\mh\lsim100\gev$.

What about other Higgs bosons associated with the $\h$.
This is a model-dependent issue.
An investigation is reported in one contribution,
\Ref\hewettetal{V. Barger, N.G. Deshpande, J.L. Hewett, and T.G. Rizzo,
{\it A Separate Higgs}, these proceedings.}
in the case where the $\h$ is required to be part
of a two-doublet model of type-I.  If the Higgs potential is to
realize the above described scenario for the lighter of the two
neutral Higgs bosons (\ie\ $\h=\hl$) naturally, then $\beta$
must be small, and the neutral sector mixing angle $\alpha$
must be near $-\pi/2$.  Thus, the couplings of the $\hh$ and $\ha$
to fermions ($\propto 1/\sinb$) will be enhanced.
These fermion couplings enter into the loop computations of
both the $gg$ and the $\gam\gam$ couplings of the $\hh$ and $\ha$.
The fermionic coupling enhancement is such, for instance, that
below $t\anti t$ threshold the $\hh$ has a $\gam\gam$ branching
ratio that generally exceeds $10^{-3}$, while the $\ha\rta\gam\gam$
branching ratio rises from $\sim 10^{-4}$ at $\mha=80\gev$
to above $10^{-3}$ for $\mha$ just below $2\mt$.

The result is that the ratio (as computed at the SSC)
$$R_{\hh,\ha}\equiv {\sigma(gg\rta \hh,\ha\rta\gam\gam)
\over \sigma(gg\rta \hsm\rta\gam\gam)}\eqn\ratio$$
is always larger than 1 for the $\hh$ (substantially so for $\mhh$
below 100 GeV), while $R_{\ha}$ can be almost 1 in the $\mha\lsim120\gev$
range.  (Both $R_{\ha}$ and $R_{\hh}$ rise to values substantially above
1 for masses above $150\gev$.)  Meanwhile, the $\ha$ and $\hh$ widths remain
well below 1 GeV for masses below 150 GeV.
Since the $\gam\gam$ inclusive mode is viable (in those detectors
with $\sim 1\%$ $\gam\gam$ mass resolution) for a $\hsm$
with $80\lsim\mhsm\lsim130\gev$, it is clear that both the $\hh$  and the
$\ha$ of this model would be detectable over this same mass
range.  Indeed, because $R_{\hh}$
becomes large for low $\mhh$, $\hh$ detection would be possible well
below the $\sim 60-80\gev$ lower limit of the inclusive
$\gam\gam$ mode that applies in the case of the $\hsm$.

The enhanced fermionic couplings of the $\hh$ and $\ha$ also imply that
the production processes $gg\rta t\anti t \hh$ and $gg\rta t\anti t \ha$
will be at least an order of magnitude larger than for the SM $\hsm$.
For $\mha$ and $\mhh$ below $2\mt$, the
above quoted $\gam\gam$ branching ratios of the $\hh$ and $\ha$ are
then such that the $\ell\gam\gam$ channel will provide signals for
the $\hh$ and $\ha$ that are at least as good as those obtained for
the SM $\hsm$, and often much better.
Thus, the $\ell\gam\gam$ and $\gam\gam$ channels {\it each}
have the potential of allowing discovery of {\it all} of the neutral
Higgs bosons of such a model.

For $\mhh$ and $\mha$ above $2\mt$, the
$\hh$ and $\ha$ decay primarily to $t\anti t$. The enhanced $t\anti t \hh$
and $t\anti t \ha$ cross sections imply
large, and probably observable,
signals for both the $\hh$ and $\ha$
in the $t\anti t t\anti t$ final state at the SSC.

In general, this model illustrates the fact that a Higgs sector
which differs significantly from those of the SM or MSSM may well
be much more easily explored.  The SM and MSSM Higgs sectors could not
have been more cunningly constructed if the goal were
to make Higgs detection at hadron colliders as challenging as possible.

\endpage

\smallskip
\noindent{\bf 5. Conclusions}
\smallskip

The techniques for probing the Higgs or EWSB sector at the SSC
continue to evolve and expand.  New modes and refinements give
us great confidence that the SSC will be a powerful tool for
revealing the Higgs bosons associated with the mechanism
for mass generation.  In addition to having demonstrated
that we can either discover the Higgs boson of the Standard Model
or detect strong $WW$ interactions if the SM Higgs is heavy, we have
now clearly established a no-lose theorem for the Higgs sector of
the Minimal Supersymmetric Model. At least one of the MSSM Higgs bosons
can be found at the SSC, regardless of model parameter choices.
These are especially important milestones given that a Higgs sector
that differs substantially from those of the SM and MSSM is often
far more easily probed, using a selection of the same production and detection
procedures as developed for the SM and MSSM.

Experimentally, two crucial items arise over and over again.  First,
it is vital that the detectors continue to stress
excellent mass resolution in the $\gam\gam$ channel and excellent
jet/photon discrimination. Secondly,
if there is a single experimental lesson
to be drawn from the most recent theoretical efforts, it is
the great importance that should be attached to the construction
of a $b$-jet vertex detector and the development of associated
$b$-identification algorithms that together yield the
highest possible efficiency and purity for $b$-tagging.

\smallskip\noindent{\bf 6. Acknowledgements} \smallskip
This review has been supported in part by Department of Energy
grant \#DE-FG03-91ER40674
and by Texas National Research Laboratory grant \#RGFY93-330.
I would like to thank the Aspen Center for Physics for support
during part of its preparation.  JFG gratefully acknowledges
the contributions of his many collaborators to the content of this
report, and the organizational efforts of L. Orr and T. Han.

\smallskip
\refout
\end